\documentclass{aastex63}

\usepackage{amssymb, amsmath}
\usepackage{xcolor}
\usepackage{rotating}
\usepackage[normalem]{ulem}
\usepackage{multirow}
\newcommand{\cmnt}[1]{}

\defcitealias{NAP12951}{(National Research Council 2010)}

\received{\today}
\submitjournal{PASP}

\shorttitle{Visible flat fields}
\shortauthors{Givans et al.}

\graphicspath{{./}{figures/}}

\begin{document}

\title{Quantum yield and charge diffusion in the {\slshape Nancy Grace Roman Space Telescope} infrared detectors}

\author[0000-0002-5870-6108]{Jahmour J. Givans}
\affiliation{Department of Astrophysical Sciences, Princeton University, 4 Ivy Lane, Princeton, New Jersey, 08540, USA}
\affiliation{Center for Computational Astrophysics, Flatiron Institute, 162 5th Ave, New York, New York, 10010}
\email{jgivans@princeton.edu}

\author[0000-0002-5636-233X]{Ami Choi}
\affiliation{Department of Physics, California Institute of Technology, 1200 E. California Blvd., Pasadena, CA 91125, USA}

\author[0000-0002-2762-2024]{Anna Porredon}
\affiliation{Center for Cosmology and AstroParticle Physics, The Ohio State University, 191 West Woodruff Avenue, Columbus, Ohio, 43210, USA}
\affiliation{Department of Physics, The Ohio State University, 191 West Woodruff Avenue, Columbus, Ohio, 43210, USA}

\author[0000-0001-6790-2939]{Jenna K.C. Freudenburg}
\affiliation{Department of Astronomy \& Astrophysics, University of Toronto, 50 St. George Street, Toronto, Ontario, M5S 3H4, Canada}
\affiliation{Canadian Institute for Theoretical Astrophysics, University of Toronto, 60 St. George Street, Toronto, Ontario, M5S 3H8, Canada}

\author[0000-0002-2951-4932]{Christopher M. Hirata}
\affiliation{Center for Cosmology and AstroParticle Physics, The Ohio State University, 191 West Woodruff Avenue, Columbus, Ohio, 43210, USA}
\affiliation{Department of Physics, The Ohio State University, 191 West Woodruff Avenue, Columbus, Ohio, 43210, USA}
\affiliation{Department of Astronomy, The Ohio State University, 140 West 18th Avenue, Columbus, Ohio, 43210, USA}

\author{Robert J. Hill}
\affiliation{Conceptual Analytics, NASA Goddard Space Flight Center, Mail Code 667, Greenbelt, Maryland, 20771, USA}

\author{Christopher Bennett}
\affiliation{Science Systems and Applications, Inc., NASA Goddard Space Flight Center, Mail Code 553, Greenbelt, Maryland, 20771, USA}

\author{Roger Foltz}
\affiliation{NASA Goddard Space Flight Center, Mail Code 553, Greenbelt, Maryland, 20771, USA}

\author{Lane Meier}
\affiliation{Conceptual Analytics, NASA Goddard Space Flight Center, Mail Code 553, Greenbelt, Maryland, 20771, USA}

\begin{abstract}
Weak gravitational lensing is a powerful tool for studying the growth of structure across cosmic time. The shear signal required for weak lensing analyses is very small, so any undesirable detector-level effects which distort astronomical images can significantly contaminate the inferred shear. The {\slshape Nancy Grace Roman Space Telescope} ({\slshape Roman}) will fly a focal plane with 18 Teledyne H4RG-10 near infrared (IR) detector arrays; these have never before been used for weak lensing and they present different instrument calibration challenges relative to CCDs. A pair of previous investigations \citep{2020PASP..132a4501H,2020PASP..132a4502C} demonstrated that spatiotemporal correlations of flat field images can effectively separate the brighter-fatter effect (BFE) and interpixel capacitance (IPC). A third paper in the series \citep{2020PASP..132g4504F} introduced a Fourier-space treatment of these correlations which allowed the authors to expand to higher orders in BFE, IPC, and classical nonlinearity (CNL). This work expands the previous formalism to include quantum yield and charge diffusion. We test the updated formalism on simulations and show that we can recover input visible characterization values to within a few percent. We then apply the formalism to visible and IR flat field data from three {\slshape Roman} flight candidate detectors. We find that BFE is present in all detectors and that the magnitude of its central pixel value is comparable between visible data and IR data. We fit a 2D Gaussian model to the charge diffusion at 0.5 $\mu$m wavelength, and find variances of $C_{11} = 0.1066\pm 0.0011$ pix$^2$ in the horizontal direction, $C_{22} = 0.1136\pm 0.0012$ pix$^2$ in the vertical direction, and a covariance of $C_{12} = 0.0001\pm 0.0007$ pix$^2$ (stat) for SCA 20829. Last, we convert the asymmetry of the charge diffusion into an equivalent shear signal using the sensitivity coefficients for the {\slshape Roman} survey, and find a contamination of the shear correlation function to be $\xi_+ \sim 10^{-6}$ for each detector. This exceeds {\slshape Roman's} allotted error budget for the measurement by a factor of $\mathcal{O}(10)$ in power (amplitude squared) but can likely be mitigated through standard methods for fitting the point spread function (PSF) since for weak lensing applications the charge diffusion can be treated as a contribution to the PSF. Further work considering the impact of charge diffusion and quantum yield on shear measurements will follow once all detectors covering the {\slshape Roman} focal plane are selected.
\end{abstract}


\section{Introduction} \label{sec:intro}

Over 20 years have passed since astronomers first obtained observational evidence for the accelerated expansion of our Universe \citep{1998AJ....116.1009R, 1999ApJ...517..565P}. When placed in context of the standard cosmological model, this evidence implies that nearly 70\% of the Universe's energy density is contained in dark energy. To better understand this mysterious component of nature, the Dark Energy Task Force \citep{2006astro.ph..9591A} recommended an extensive observational program for studying cosmic expansion history and the growth of structure. Included in this endeavor are observations of baryon acoustic oscillations, galaxy clusters, Type Ia supernovae, and weak gravitational lensing (see, e.g., \citealt{2013PhR...530...87W} for an overview). Weak gravitational lensing induces correlations between the apparent shapes of distant galaxies; its statistical properties are a powerful tool for precision cosmology because they are sensitive both to the structures in the Universe as well as the background geometry (expansion history) of the Universe. These correlations in weak gravitational lensing are small and thus challenging to observe from both the point of view of sample size and systematic error control. The early detections of weak lensing $\sim 20$ years ago \citep{2000A&A...358...30V, 2000MNRAS.318..625B, 2000Natur.405..143W, 2000astro.ph..3338K} have since given way to larger, more ambitious survey programs such as the Hyper Suprime Cam \citep{2019PASJ...71...43H, 2020PASJ...72...16H}, the Kilo Degree Survey \citep{2020A&A...633A..69H, 2020A&A...634A.127A}, and the Dark Energy Survey \citep{2018PhRvD..98d3528T, 2021arXiv210513543A,2021arXiv210513544S}.

The {\slshape Nancy Grace Roman Space Telescope} ({\slshape Roman}) was recommended in the Astro2010 Decadal Survey \citetalias{NAP12951}, and underwent several design iterations before being approved \citep{2012arXiv1208.4012G, 2013arXiv1305.5422S, 2015arXiv150303757S, 2019arXiv190205569A}. {\slshape Roman} is a near-infrared (NIR) survey telescope that will carry out imaging and spectroscopic surveys for the aforementioned dark energy probes, a time-domain survey of the Galactic Bulge to study exoplanet demographics via microlensing, and other astronomical observations that benefit from NIR survey coverage or space-based resolution and stability; and it will carry a coronagraph as a technology demonstration for direct imaging of exoplanets. {\slshape Roman} will have a wide field of view (0.281~deg$^2$), and the Reference Survey proposes to observe 2000~deg$^2$ in four bands from 0.9--2.0 $\mu$m at a depth of 26.2--26.95 mag AB (at 5$\sigma$). 

Like other weak lensing surveys, {\slshape Roman} will use a mosaic camera --- in this case, composed of 18 detector arrays with $4088\times 4088$ active pixels each for a total of 300 Mpix. However, most weak lensing surveys --- including the other large planned surveys with {\slshape Euclid}\footnote{{\slshape Euclid} carries a NIR focal plane with 16 H2RG detector arrays. However shape measurement on {\slshape Euclid} is planned to be carried out with the better-sampled visible CCD channel.} \citep{2011arXiv1110.3193L} and the Vera Rubin Observatory \citep{2019ApJ...873..111I} --- operate in the visible part of the spectrum with silicon charge-coupled devices (CCDs). Recognizing the need for robust wavelength coverage across the visible and NIR bands for photometric redshifts, as well as the intense airglow that dominates ground-based NIR imaging, Astro2010 recommended a split strategy including a large visible survey from the ground and NIR observations from space. The NIR also offers several advantages for galaxy shape measurement: the wavefront stability requirements are less stringent as one goes to longer wavelengths \citep{2021MNRAS.501.2044T}; and for typical galaxy colors and space backgrounds the achievable signal-to-noise ratio is higher. However, the NIR detector architecture is very different from a CCD; the technology for large-format arrays matured more recently \citep{2014SPIE.9154E..2HP}, and weak lensing analysts must learn the particular ``features'' of these detectors in order to derive robust constraints. Of particular interest to us is understanding how {\slshape Roman} detector systematics will impact the weak lensing shear autocorrelation signal. Making accurate measurements of this small signal requires controlling all sources of galaxy shape measurement systematics, including those originating in detectors, to a level of a few$\times 10^{-4}$.

The {\slshape Roman} H4RG NIR detector arrays \citep{2020JATIS...6d6001M} are part of the Teledyne HxRG family \citep{2011ASPC..437..383B}, although the physical pixel pitch of 10 $\mu$m is smaller than the 18 $\mu$m used in the H1RG ({\slshape Hubble Space Telescope} Wide Field Camera 3) and H2RG ({\slshape James Webb Space Telescope}, {\slshape Euclid}, SPHEREx) detector arrays. Light passes through an anti-reflection coating and is converted to electron-hole pairs in a mercury cadmium telluride (HgCdTe, 2.5 $\mu$m cutoff) layer. Each pixel is formed from a p-n junction in the HgCdTe; holes are collected on the p-type side and the voltage difference across the junction decreases as the pixel is exposed to light. An indium interconnect in each pixel connects the p-type side of the diode to a readout integrated circuit (ROIC), which is used to address each pixel in sequence and non-destructively read its voltage. This architecture leads to some effects that are similar to CCDs, and some that are very different.

In order to prepare for weak lensing with {\slshape Roman}, previous studies have both simulated the impacts of specific detector effects, as well as used laboratory studies to assess their importance. Interpixel capacitance (IPC), which is inevitable because the pixels have conducting interconnects in close proximity when they are read out, causes charge in one pixel to induce a signal in its neighbors \citep{2004SPIE.5167..204M}. Since the pixel boundaries are defined by electric fields, the ``brighter-fatter effect (BFE)'' seen in CCDs --- where a pixel containing more charge alters the electric field geometry and subsequent charges are less likely to be collected in that pixel \citep[e.g.,][]{2014JInst...9C3048A, 2015JInst..10C5032G} --- also occurs in NIR detectors, although it is shorter range due to the relatively thin active layer \citep{2020PASP..132a4502C}. NIR detectors also exhibit ``classical'' non-linearity (CNL), a non-linearity of the mapping from collected charge to digitized signal that is typically modeled as a polynomial in modern pipelines \citep[e.g.,][]{2014wfc..rept...17H}. Charge traps in the HgCdTe lead to additional time-dependent effects such as persistence (an after-image of bright sources in subsequent images) and count rate-dependent non-linearity \citep{2008SPIE.7021E..0JS}. The charge trapping effects are considered in ongoing work and will not be addressed in this paper. The impact of IPC on weak lensing is summarized in \citet{2016PASP..128i5001K}; that of CNL in \citet{2016PASP..128j4001P}; and that of BFE in \citet{2018PASP..130f5004P}.

Previous investigations have used Poisson fluctuations in laboratory flat fields to characterize CNL, IPC, and BFE. In CCDs, correlations between neighboring pixels in a flat field contain a wealth of information about the BFE \citep{2014JInst...9C3048A, 2015A&A...575A..41G, 2019A&A...629A..36A}. However, for NIR detectors there are also correlations induced by IPC \citep{2004SPIE.5167..204M, 2006OptEn..45g6402M}, and it is necessary to solve for both effects. \citet{2020PASP..132a4501H} took advantage of the non-destructive readout capabilities of HxRG-10s to separate BFE from IPC and model their interaction with CNL. Their work also presented \textsc{Solid-waffle}, a code used to simulate flats and analyze correlations in real HxRG flats to extract BFE and IPC parameters. The authors found a $\sim 12\%$ lingering bias in the BFE kernel when their formalism was applied to simulated data. \citet{2020PASP..132a4502C} applied the same formalism to flats measured on {\slshape Roman} development detector sensor chip assembly (SCA) 18237 and detected a correlation signal after correction for IPC; they presented several lines of evidence that this correlation was due primarily to the BFE.
In \citet{2020PASP..132g4504F} the authors reworked the formalism of \citet{2020PASP..132a4501H} in Fourier space which allowed them to retain higher-order nonlinearity terms that were previously discarded. After incorporating these changes in \textsc{Solid-waffle}, the authors tested the updated formalism on simulated flats and found that the BFE kernel bias was reduced to $<1\%$. They also applied the formalism to
SCA 18237 and flight candidate detectors SCA 20663, SCA 20828, and SCA 20829; detailed results from those analyses are summarized in \citet[\S6]{2020PASP..132g4504F}.

In our previous studies, we assumed that a photon interacted with the absorbing semiconductor layer and generated a single electron-hole pair.\footnote{It is actually the hole that is collected in a pixel, although for historical reasons we refer to collected charge in ``electrons'' and current in ``electrons per second.''} It is possible, however, for a photon of sufficiently high energy (short wavelength) to interact with a detector and produce two electron-hole pairs\footnote{In principle, a photon of even higher energy can generate $>2$ electron-hole pairs. The techniques in this paper could be extended to investigate those scenarios, but constraining the probability for every number of electron-hole pairs would require going beyond 2-point statistics and is beyond the scope of our work.}; the threshold for this behavior must be at least twice the band gap energy, and in studies with James Webb Space Telescope NIRSpec arrays (HgCdTe, 5 $\mu$m cutoff) it was $\sim 2.65E_{\rm gap}$ \citep{2014PASP..126..739R}. This has several effects on flat field correlations, which are both a complication and an opportunity \citep{2008PASP..120..759M}. There is an additional contribution to the variance of the flat field data since the process of generating 1 or 2 pairs is stochastic (Fano noise: \citealt{1947PhRv...72...26F}). Such occurrences increase the {\slshape Roman} detector array's quantum yield beyond expectations for IR-only absorptions. Our previous work also ignored the details of charge diffusion, the process by which a hole must travel through an absorber before reaching depleted HgCdTe en route to a pixel, treating it as ``direct walk'' between points A and B. Charge diffusion in H4RG-10 detectors is more complex than this and is a source of systematics for weak lensing measurements \citep{2020JATIS...6d6001M}.

The purpose of this work is to repeat the analysis of \citet{2020PASP..132g4504F} with updates to the formalism and \textsc{Solid-waffle} to include quantum yield and charge diffusion effects. This paper is organized as follows: In \S\ref{sec:formalism} we present our formalism
for describing BFE, IPC, CNL, quantum yield, and charge diffusion treated as independent effects. We follow with a more detailed discussion of Gaussian charge diffusion in \S\ref{sec:Gauss}. \S\ref{sec:ipnl_corr} builds the mathematical framework for describing all five detector effects acting in concert. In \S\ref{sec:data} we describe the flat and dark data along with our simulations. In \S\ref{sec:analysis} we explain how {\sc Solid-waffle} computes visible characterization parameters and test this against simulations. We then analyze real data from three {\slshape Roman} flight candidates in \S\ref{sec:results}. We conclude with a discussion of our findings and future plans in \S\ref{sec:discussion}.

\section{Formalism} \label{sec:formalism}

We follow closely the formalism of \citet{2020PASP..132g4504F} --- itself based on \citet{2020PASP..132a4501H} --- but with some updates to account for quantum yield effects.

As in \citet{2020PASP..132g4504F}, we define Fourier transforms and inverse Fourier transforms of a quantity $F$ according to
\begin{equation}
    \widetilde{F}(k_1,k_2)=\sum_{x_1=0}^{N-1}\sum_{ x_2=0}^{N-1}F(x_1,x_2) e^{-2\pi i(k_{1} x_1+k_{2} x_{2})/N}
~~~\leftrightarrow ~~~
    {F}(x_1,x_2)=\frac{1}{N^2}\sum_{k_1=0}^{N-1}\sum_{ k_2=0}^{N-1}\widetilde{F}(k_1,k_2) e^{2\pi i(k_{1} x_1+k_{2} x_{2})/N},
\end{equation}
where we consider an $N\times N$ region of pixels with periodic boundary conditions (of course a real detector does not have periodic boundary conditions, but if $N$ is taken to be large enough then for effects with range of up to a few pixels, such as IPC, BFE, and charge diffusion, the nature of the boundary conditions is not important and we make the simplest choice). Pixel indices are taken to be 2-component vectors with integer coordinates; $x_1$ is the column number and $x_2$ is the row number. The Fourier domain indices $k_1$ and $k_2$ are also integers; the wave vector is $(u_1,u_2) = (k_1/N,k_2/N)$ cycles per pixel, and for large $N$ is in the square domain $-\frac12 < u_1,u_2 < \frac12$.

In \citet{2020PASP..132g4504F}, ``$I$'' represented the current per pixel (electrons per pixel per second, or e/p/s). In this paper, where one photon may generate 2 electron-hole pairs, we distinguish the electrical current $I_e$ (e/p/s) from the number of photons absorbed per pixel per second, $I_\gamma$ (photons/p/s). Note that the incident photon flux is $>I_\gamma$, because some photons do not make it through the anti-reflection coating or otherwise do not result in collected charge. If a photon has a probability $1-\omega$ of generating 1 collected carrier, and probability $\omega$ of generating 2 collected carriers, then
\begin{equation}
I_e = (1-\omega)( I_\gamma) +( \omega)( 2I_\gamma ) = (1+\omega)I_\gamma.
\label{eq:IeIg}
\end{equation}
The factor $1+\omega$ is the quantum yield, often denoted by $\eta$.

We describe the statistics of collecting 2 carriers as follows. Suppose that a single photon produces two identical carriers that are collected, A and B. If A is collected at pixel $(x_1,x_2)$, then the conditional probability of B being collected in $(x_1+\Delta x_1,x_2+\Delta x_2)$ is $p_2(\Delta x_1,\Delta x_2)$. We take this to depend only on $\Delta x_1$ and $\Delta x_2$, taking into consideration the assumed discrete translation symmetry within a patch of pixels. Note that probabilities sum to 1: $\sum_{\Delta x_1,\Delta x_2} p_2(\Delta x_1,\Delta x_2) = 1$. Also since A and B are identical, we have by construction $p_2(\Delta x_1,\Delta x_2) = p_2(-\Delta x_1,-\Delta x_2)$. In the absence of charge diffusion, both A and B would always land in the same pixel, and we would have $p_2(\Delta x_1,\Delta x_2) = \delta_{\Delta x_1,0} \delta_{\Delta x_2,0}$. As charge diffusion increases, the $p_2$ kernel spreads out.

For convenience later, we define the combined kernel
\begin{equation}
\Phi(\Delta x_1,\Delta x_2) = \frac{\omega}{1+\omega} p_2(\Delta x_1, \Delta x_2).
\label{eq:Phi-def}
\end{equation}
This contains the same information as $\omega$ and $p_2$ separately (since $p_2$ is normalized: $\sum_{\Delta{\boldsymbol x}} p_2(\Delta{\boldsymbol x})=1$), but it is well-behaved in the limit $\omega\rightarrow 0$ where $p_2$ becomes undefined.

As in our previous papers, we include CNL and IPC as follows. The change in signal in a pixel is
\begin{equation}
S_{\rm initial}(x_1,x_2) - S_{\rm final}(x_1,x_2) = \frac1g
\left[ Q_{\rm conv}(x_1,x_2)
- \sum_{\nu = 2}^n \beta_\nu (Q_{\rm conv}(x_1,x_2))^\nu
\right],
\label{eq:K1}
\end{equation}
where $S(x_1,x_2)$ is the signal in data numbers (DN) in pixel $(x_1,x_2)$; $g$ is the conversion gain (units: e/DN); $\beta_\nu$ are the coefficients of a classical non-linearity polynomial fit of degree $n$; and the charge after IPC convolution is
\begin{equation}
Q_{\rm conv}(x_1,x_2)
=
\sum_{\Delta x_1,\Delta x_2} [K_{\Delta x_1,\Delta x_2} + K^I_{\Delta x_1,\Delta x_2}\bar Q]Q_{x_1-\Delta x_1, x_2-\Delta x_2}.
\label{eq:K2}
\end{equation}
Here $Q(x_1,x_2)$ is the charge in pixel $(x_1,x_2)$; $K$ is the IPC kernel; and $K^I$ is an IPC non-linearity that allows the IPC to vary with flat field level $\bar Q$. (More general types of IPC non-linearity are possible when pixels are illuminated to very different levels --- see, e.g., \citealt{2016SPIE.9915E..2ID, 2017OptEn..56b4103D, 2018PASP..130g4503D} --- but flat field correlations are sensitive only to $K^I$.) The kernel is normalized to sum to unity, $\sum_{\Delta x_1,\Delta x_2} K_{\Delta x_1,\Delta x_2} = 1$. The ``nearest-neighbor'' IPC kernel coefficients, $K_{\pm 1,0}$ and $K_{0,\pm 1}$, are denoted by $\alpha$ (and $\alpha_{\rm H}$ or $\alpha_{\rm V}$ if the horizontal or vertical neighbors are different). True electrostatic IPC is symmetric: $K_{\Delta x_1,\Delta x_2} = K_{-\Delta x_1,-\Delta x_2}$, although other forms of cross-talk that may mimic IPC are not necessarily symmetric. The largest of these is the vertical trailing pixel effect (VTPE), described at length in \citet{2020PASP..132g4504F}, although we will not explicitly model it here. The description in Eqs.~(\ref{eq:K1},\ref{eq:K2}) performs the IPC correction before the CNL; in a physical detector, the two effects are mixed and do not have a well-defined ordering, but for linearized fluctuations around a flat field (i.e., where we can take the first-order Taylor expansion of the CNL polynomial around the mean accumulated charge $\bar Q$), the mathematical operations commute and the issue is moot.

The BFE is modeled using the \citet{2014JInst...9C3048A} model, where the effective area of a pixel ${\cal A}_{x_1,x_2}$ is modified by the charge in the pixels near it:
\begin{equation}
{\cal A}_{x_1,x_2} = {\cal A}_{x_1,x_2}^0 \left[ 1 +
\sum_{\Delta x_1,\Delta x_2} a_{\Delta x_1,\Delta x_2} Q(x_1+\Delta x_1,x_2+\Delta x_2)
\right].
\end{equation}
The coefficients $a_{\Delta x_1,\Delta x_2}$ form the BFE kernel. The BFE could be wavelength-dependent because shorter wavelengths of light have a smaller mean free path in HgCdTe, and hence electron-hole pairs are produced closer to the illuminated surface of the detector. Therefore in what follows, when we use two wavelengths of light (1.4 $\mu$m and 0.5 $\mu$m), we fit the BFE coefficients again each time we change the wavelength.

Both BFE and non-linear IPC are examples of inter-pixel non-linearity (IPNL). The combination $K^2a+KK^I$ (where combinations of kernels such as $K^2a$ indicate convolution) appears commonly in correlation function analyses and will be called the ``IPNL kernel.'' \citet{2020PASP..132a4502C} showed that for the {\slshape Roman} detectors, BFE is the dominant form of IPNL.

All lateral lengths in this paper are in units of pixels (10 $\mu$m for {\slshape Roman}) unless otherwise specified.

\section{Pairwise pixel probability for Gaussian charge diffusion}
\label{sec:Gauss}

We now consider $p_2(\Delta x_1,\Delta x_2)$ and its Fourier transform for the special case of Gaussian charge diffusion. The technique of this section generalizes easily to other charge diffusion probability densities, but not to cases with intra-pixel sensitivity variation.

A charge generated at position ${\boldsymbol X}$ diffuses to a location ${\boldsymbol X} + {\boldsymbol\xi}$ with lateral displacement ${\boldsymbol\xi}$ and 2D probability density $P_{\rm cd}({\boldsymbol\xi})$. In the Gaussian case, this is a 2D normalized Gaussian,
\begin{equation}
    P_{\rm cd}(\xi_1,\xi_2) = {\mathcal N}_{\bf C}(\xi_1,\xi_2) = \frac1{2\pi\sqrt{C_{11}C_{22}-C_{12}^2}}\exp \left[ -\frac{C_{22}\xi_1^2+C_{11}\xi_2^2 - 2C_{12}\xi_1\xi_2}{2(C_{11}C_{22}-C_{12}^2)} \right],
\end{equation}
where ${\bf C}$ is a $2\times 2$ symmetric covariance matrix, and ${\mathcal N}_{\bf C}$ represents the Gaussian distribution centered on 0 and with covariance ${\bf C}$. We further allow for a pixel to have a top-hat function $W_{\rm p}({\boldsymbol y})$ given by
\begin{equation}
W_{\rm p}(y_1,y_2) = \left\{ \begin{array}{lll}
1 & & |y_1|<\frac12 ~~{\rm and}~~ |y_2|<\frac12 \\ 0 & & {\rm otherwise} \end{array} \right.~~.
\end{equation}

We suppose that a photon generates two carriers at position ${\boldsymbol X}$. Now the probability for carrier B to land in pixel ${\boldsymbol x}+\Delta{\boldsymbol x} = (x_1 + \Delta x_1, x_2 + \Delta x_2)$ given that carrier A landed in pixel ${\boldsymbol x} = (x_1,x_2)$ is given by Bayes's theorem:
\begin{equation}
p_2(\Delta{\boldsymbol x}) = P({\rm B~in~}{\boldsymbol x}+\Delta {\boldsymbol x} | {\rm A~in~}{\boldsymbol x})
= \frac{P({\rm B~in~}{\boldsymbol x}+\Delta {\boldsymbol x} ~{\rm and}~ {\rm A~in~}{\boldsymbol x})}{P({\rm A~in~}{\boldsymbol x})}.
\end{equation}
Writing these probabilities as integrals over ${\boldsymbol X}$ and the 2D diffusion vectors ${\boldsymbol\xi}_{\rm A}$ and ${\boldsymbol\xi}_{\rm B}$, and dropping the normalizing factor that cancels out, gives
\begin{equation}
p_2(\Delta{\boldsymbol x}) = 
\frac{\int d^2{\boldsymbol X}
\int d^2{\boldsymbol \xi}_{\rm A}
\int d^2{\boldsymbol \xi}_{\rm B}
~ P_{\rm cd}({\boldsymbol \xi}_{\rm A})
P_{\rm cd}({\boldsymbol \xi}_{\rm B})
W_{\rm p}({\boldsymbol X}+{\boldsymbol \xi}_{\rm A}-{\boldsymbol x})
W_{\rm p}({\boldsymbol X}+{\boldsymbol \xi}_{\rm B}-{\boldsymbol x}-\Delta{\boldsymbol x})}{
\int d^2{\boldsymbol X}\int d^2{\boldsymbol \xi}_{\rm A}
~ P_{\rm cd}({\boldsymbol \xi}_{\rm A})
W_{\rm p}({\boldsymbol X}+{\boldsymbol \xi}_{\rm A}-{\boldsymbol x})
}.
\end{equation}
The denominator integrates to 1 (to see this, do the $d^2{\boldsymbol X}$ integral first to get $\int d^2{\boldsymbol \xi}_{\rm A}
~ P_{\rm cd}({\boldsymbol \xi}_{\rm A})$, which is the normalization integral for $P_{\rm cd}$). For the numerator, we can change variables to ${\boldsymbol\eta} = {\boldsymbol\xi}_{\rm B} - {\boldsymbol\xi}_{\rm A}$ and ${\boldsymbol y} = {\boldsymbol X}+{\boldsymbol \xi}_{\rm A}-{\boldsymbol x}$, and get
\begin{equation}
p_2(\Delta{\boldsymbol x}) = 
\int d^2{\boldsymbol y}
\int d^2{\boldsymbol \xi}_{\rm A}
\int d^2{\boldsymbol \eta}
~ P_{\rm cd}({\boldsymbol \xi}_{\rm A})
P_{\rm cd}({\boldsymbol\eta} + {\boldsymbol \xi}_{\rm A})
W_{\rm p}({\boldsymbol y})
W_{\rm p}({\boldsymbol y} + {\boldsymbol\eta}-\Delta{\boldsymbol x}).
\end{equation}
The $d^2{\boldsymbol y}$ and $d^2{\boldsymbol\xi_{\rm A}}$ integrals are now convolution integrals that can be done analytically, leaving us with
\begin{equation}
p_2(\Delta{\boldsymbol x}) = \int d^2{\boldsymbol \eta}~
{\mathcal N}_{2{\bf C}}({\boldsymbol\eta}) T_{\rm p}({\boldsymbol \eta}-\Delta{\boldsymbol x})
= \int d^2{\boldsymbol \eta}~
{\mathcal N}_{2{\bf C}}({\boldsymbol z} + \Delta{\boldsymbol x}) T_{\rm p}({\boldsymbol z})
,
\label{eq:p2int}
\end{equation}
where ${\mathcal N}_{2{\boldsymbol C}}({\boldsymbol\eta})$ is the Gaussian distribution with covariance $2{\bf C}$ and
\begin{equation}
T_{\rm p}(z_1,z_2) =
\int d^2{\boldsymbol y}~
 W_{\rm p}({\boldsymbol y})
W_{\rm p}({\boldsymbol y} + {\boldsymbol z}) =
\left\{ \begin{array}{lll}
(1-|z_1|)(1-|z_2|) & & |z_1|<1 ~~{\rm and}~~ |z_2|<1 \\ 0 & & {\rm otherwise} \end{array} \right.
\end{equation}
is the autocorrelation of the tophat (a 2D triangle function). In the second part of Eq.~(\ref{eq:p2int}), we have substituted ${\boldsymbol z} = {\boldsymbol\eta}-\Delta{\boldsymbol x}$.

Equation~(\ref{eq:p2int}) can be evaluated numerically. The integrand is only non-zero in the square domain where $-1< z_1<1$ and $-1<z_2<1$, and for analytic evaluation purposes this breaks down into 4 sub-squares that can be evaluated separately. The integration becomes more difficult for small charge diffusion, since the integrand must then be more finely sampled. We perform each of these integrations with a 2D version of Simpson's extended rule (Eq.~4.1.14 of \citealt{1992nrca.book.....P}, implemented in both $z_1$ and $z_2$, with $N=256$), which achieves $\sim 10^{-9}$ accuracy even with an unrealistically low charge diffusion of $\sigma_{\rm cd}=0.1$ pixels (rms on both the $x_1$ and $x_2$ axes).

\section{Correlation functions with interpixel nonlinearity}\label{sec:ipnl_corr}
We now turn our attention to creating a model which describes correlations measured on {\slshape Roman} detectors in the presence of interpixel nonlinearity. This was done in real space \citep{2020PASP..132a4501H} and Fourier space \citep{2020PASP..132g4504F} for the case where one photon generates one charge carrier; the following formalism is constructed to handle the case where a single photon generates two charge carriers. Calculations presented below parallel those given in Section 3 of \citet{2020PASP..132g4504F}. We will reference those calculations as needed in this work using the format ``F/X'' which is understood to mean ``Equation X in \citet{2020PASP..132g4504F}.''

The two-point function for charge in modes $(k_1,k_2)$ and $(k_1',k_2')$ at time $t+\delta t$ given the states $\widetilde{Q}(k_1,k_2,t)$ and $\widetilde{Q}(k_1',k_2',t)$ at time $t$ is (F/16)
\begin{equation}\label{eqn:secondmoment}
 \begin{aligned}
    \langle\widetilde{Q}(k_1,k_2,&t+\delta t)\widetilde{Q}(k_1',k_2',t+\delta t)\rangle|_t =  
    \widetilde{Q}(k_1,k_2,t)\widetilde{Q}(k_1',k_2',t)
    + I_e\widetilde W(k_1,k_2,t)\widetilde{Q}(k_1',k_2',t)\delta t \\
    &+  I_e\widetilde W(k_1',k_2',t)\widetilde{Q}(k_1,k_2,t)\delta t
    + \mathrm{Cov}[\Delta\widetilde Q (k_1,k_2,t),\Delta\widetilde Q (k_1',k_2',t)],
 \end{aligned}
\end{equation}
where the last term is equivalent to $\langle \Delta\widetilde Q (k_1,k_2,t)\Delta\widetilde Q (k_1',k_2',t)\rangle$ after dropping the term of order $\delta t^2$. The configuration-space covariance term is 

\begin{equation}
 \begin{aligned}
    \mathrm{Cov}[\Delta Q (x_1,x_2,t),\Delta Q (x_1',x_2',t)] = (1+\omega)I_{\gamma}\langle W(x_1,x_2;t)\rangle\delta_{x_1,x_1'}\delta_{x_2,x_2'}\delta t +2\omega I_{\gamma}\langle W(x_1,x_2;t)\rangle p_2(x_1'-x_1,x_2'-x_2)\delta t,
 \end{aligned}
\end{equation}
which can be expressed in Fourier space as
\begin{equation}
    \mathrm{Cov}[\Delta\tilde Q (k_1,k_2,t),\Delta\tilde Q (k_1',k_2',t)] = (1+\omega)I_{\gamma}\langle\widetilde{W}(k_1+k_1',k_2+k_2';t)\rangle\delta t + 2\omega I_{\gamma}\langle\widetilde{W}(k_1+k_1',k_2+k_2';t)\rangle\tilde{p}_2(k_1',k_2')\delta t.
\label{eq:CovQQ}
\end{equation}

Take the expectation value of Equation (\ref{eqn:secondmoment}) in the limit $\delta t \rightarrow 0$ and rearrange the result to obtain
\begin{equation}\label{eq:ODE}
    \begin{aligned}
    \frac{d}{dt} \langle\widetilde{Q}(k_1,k_2,t)\widetilde{Q}(k_1',k_2',t)\rangle &= I_e[\tilde{a}^{*}(k_1,k_2)+\tilde{a}^{*}(k_1',k_2')]\langle\widetilde{Q}(k_1,k_2,t)\widetilde{Q}(k_1',k_2',t)\rangle \\ &+ I_e N^2[\delta_{k_1,k_2,0}\langle\widetilde{Q}(k_1',k_2',t)\rangle + \delta_{k_1',k_2',0}\langle\widetilde{Q}(k_1,k_2,t)\rangle] \\ &+ I_{\gamma}[(1+\omega)+2\omega\tilde{p}_2(k_1',k_2')][N^2 \delta_{k_1+k_1',k_2+k_2',0} + \tilde{a}^*(k_1+k_1',k_2+k_2')\langle\widetilde{Q}(k_1+k_1',k_2+k_2',t)\rangle]
    \end{aligned}
\end{equation}
The above result can be simplified using F/15 which states
\begin{equation}
    \langle \widetilde Q(k_1,k_2,t) \rangle = \frac{N^2}{\widetilde{a}^*(k_1,k_2)} \left(e^{I_e\widetilde{a}^*(k_1,k_2)t}-1\right)\delta_{k_1,0}\delta_{k_2,0}
\end{equation}
and making a few notational simplifications. $^{*}$ denotes a complex conjugate, and we refer to $\widetilde{a}^{*}(k_1',k_2')$ as $\widetilde{a}^{*\prime}$ and $\widetilde{a}^{*}(k_1+k_1',k_2+k_2')$ as $\widetilde{a}^{*+}$. Similarly we write $\widetilde{Q}(k_1',k_2',t)$ as $\widetilde{Q}'(t)$. We make the definitions
\begin{equation}\label{eqn:deltas}
\delta_{k,k',0}\equiv\delta_{k_1,0}\delta_{k_2,0}\delta_{k_1',0}\delta_{k_2',0} ~~~{\rm and}~~~
\delta_{k+k',0}\equiv\delta_{k_1+k_1',0}\delta_{k_2+k_2',0}.
\end{equation}
With these changes, the solution to Equation (\ref{eq:ODE}) with initial condition $\langle\widetilde{Q}(t=0)\widetilde{Q}'(t=0)\rangle = 0$ is
\begin{equation}\label{eq:equal-time}
    \begin{aligned}
    \langle \widetilde{Q}(t)\widetilde{Q}'(t)\rangle &= \frac{N^4\delta_{k,k',0}}{\widetilde{a}^*\widetilde{a}^{*\prime}}\Big(e^{I_e\widetilde{a}^*t}-1\Big)\left(e^{I_e\widetilde{a}^{*\prime}t} - 1\right) + \frac{N^2\delta_{k+k',0}[2\omega\tilde{p}_2(k')+(1+\omega)]}{(1+\omega)(\widetilde{a}^*+\widetilde{a}^{*\prime}-\widetilde{a}^{*+})}\Big(e^{I_e\left(\widetilde{a}^*+\widetilde{a}^{*\prime}\right)t}-e^{I_e\widetilde{a}^{*+}t}\Big).
    \end{aligned}
\end{equation}

Having solved the equal-time correlation function, we now move on to solving the unequal-time correlation function. Start with F/12 and multiply by $\tilde{Q}'(t_1)$ to obtain F/22:
\begin{equation}
    \langle\widetilde{Q}(k_1,k_2,t+\delta t)\widetilde{Q}(k_1',k_2',t_1)\rangle|_{t,t_1} =  \widetilde Q(k_1,k_2,t)\widetilde{Q}(k_1',k_2',t_1) + I_e\widetilde W(k_1,k_2,t)\widetilde{Q}(k_1',k_2',t_1)\delta t.
\end{equation}
Take the expectation value in the limit $\delta t \rightarrow 0$ to get F/23:
\begin{equation}
    \frac{d}{dt}\langle \widetilde{Q}(t)\widetilde{Q}^{\prime}(t_1)\rangle = \frac{I_e N^4\delta_{k,k',0}}{\tilde{a}^{*\prime}}\Big(e^{I_e\widetilde{a}^*t_1}-1\Big)+I_e \tilde{a}^* \langle \widetilde{Q}(t)\widetilde{Q}^{\prime}(t_1)\rangle.
\end{equation}
At $t=t_1$ the solution to this differential equation must reduce to the equal-time result of Equation (\ref{eq:equal-time}). The full solution is therefore
\begin{equation}
    \begin{aligned}
    \langle\widetilde{Q}(t)\widetilde{Q}'(t_1)\rangle &= \frac{N^4\delta_{k,k',0}}{\widetilde{a}^*\widetilde{a}^{*\prime}}\Big(e^{I_e\widetilde{a}^{*\prime}t_1}-1\Big)\Big(e^{I_e\widetilde{a}^*t}-1\Big) \\ &+ \frac{N^2\delta_{k+k',0}[2\omega\tilde{p}_2(k')+(1+\omega)]}{(1+\omega)(\widetilde{a}^*+\widetilde{a}^{*\prime}-\widetilde{a}^{*+})}\;e^{I_e\widetilde{a}^*(t-t_1)}\left(e^{I_e\left(\widetilde{a}^*+\widetilde{a}^{*\prime}\right)t_1}-e^{I_e\widetilde{a}^{*+}t_1}\right).
    \end{aligned}
\end{equation}
This is written suggestively to make clear that the first term is $\langle\widetilde{Q}(t)\rangle\times\langle\widetilde{Q}'(t_1)\rangle$, which means the covariance term is
\begin{equation}\label{eq:unequal-time}
    \textrm{Cov}[\widetilde{Q}(t),\widetilde{Q}'(t_1)] = \frac{N^2\delta_{k+k',0}[2\omega\tilde{p}_2(k')+(1+\omega)]}{(1+\omega)(\widetilde{a}^*+\widetilde{a}^{*\prime}-\widetilde{a}^{*+})}\;e^{I_e\widetilde{a}^*(t-t_1)}\left(e^{I_e\left(\widetilde{a}^*+\widetilde{a}^{*\prime}\right)t_1}-e^{I_e\widetilde{a}^{*+}t_1}\right).
\end{equation}

We are now in a position to evaluate the signal power spectrum across times $(t_a,t_b,t_c,t_d)$ which is defined as
\begin{equation}
    \widetilde{C}_{abcd}(k_1'-k_1,k_2'-k_2) = \mathrm{Cov}\left[\widetilde{S}_a(k_1,k_2)-\widetilde{S}_b(k_1,k_2),\widetilde{S}_c(k_1',k_2')-\widetilde{S}_d(k_1',k_2')\right],
\end{equation}
where $\widetilde{S}_0 - \widetilde{S}_a = f(\widetilde{Q}_a)$ is the drop in signal level from initial time $t_0$ to final time $t_a$. The form of $f(\widetilde{Q}_a)$ depends on which detector effects are considered. These are given explicitly in \citet{2020PASP..132g4504F} for BFE (F/28), IPC (F/29), IPC + NL-IPC (F/30), and CNL (F/34). The full power spectrum combining all these effects is (F/36):
\begin{equation}\label{eqn:fullcorr}
\begin{aligned}
\widetilde{C}_{abcd}^{\textrm{full}}
&= \frac{1}{g^{2}}\left[\bigg(1-\sum_{\nu=2}^n\nu\beta_\nu\overline{Q}_a^{\nu-1}\bigg)\bigg(1-\sum_{\nu=2}^n\nu\beta_\nu\overline{Q}_c^{\nu-1}\bigg)\left(\widetilde{K}+\widetilde{K}^{I}\overline{Q}_{a}\right)\left(\widetilde{K}'+\widetilde{K}^{I\prime}\overline{Q}_{c}\right)\textrm{Cov}\left(\widetilde{Q}_{a},\widetilde{Q}'_{c}\right)\right. \\ 
&\hspace{20pt}-\bigg(1-\sum_{\nu=2}^n\nu\beta_\nu\overline{Q}_a^{\nu-1}\bigg)\bigg(1-\sum_{\nu=2}^n\nu\beta_\nu\overline{Q}_d^{\nu-1}\bigg)\left(\widetilde{K}+\widetilde{K}^{I}\overline{Q}_{a}\right)\left(\widetilde{K}'+\widetilde{K}^{I\prime}\overline{Q}_{d}\right)\textrm{Cov}\left(\widetilde{Q}_{a},\widetilde{Q}'_{d}\right) \\ 
&\hspace{20pt}-\bigg(1-\sum_{\nu=2}^n\nu\beta_\nu\overline{Q}_b^{\nu-1}\bigg)\bigg(1-\sum_{\nu=2}^n\nu\beta_\nu\overline{Q}_c^{\nu-1}\bigg)\left(\widetilde{K}+\widetilde{K}^{I}\overline{Q}_{b}\right)\left(\widetilde{K}'+\widetilde{K}^{I\prime}\overline{Q}_{c}\right)\textrm{Cov}\left(\widetilde{Q}_{b},\widetilde{Q}'_{c}\right) \\ 
&\hspace{20pt}+\left.\bigg(1-\sum_{\nu=2}^n\nu\beta_\nu\overline{Q}_b^{\nu-1}\bigg)\bigg(1-\sum_{\nu=2}^n\nu\beta_\nu\overline{Q}_d^{\nu-1}\bigg)\left(\widetilde{K}+\widetilde{K}^{I}\overline{Q}_{b}\right)\left(\widetilde{K}'+\widetilde{K}^{I\prime}\overline{Q}_{d}\right)\textrm{Cov}\left(\widetilde{Q}_{b},\widetilde{Q}'_{d}\right)\right],
\end{aligned}
\end{equation}
where the covariance terms are given by Equation (\ref{eq:unequal-time}).

\section{Data}\label{sec:data}

\subsection{Flat and dark data} \label{ss:fd}

\begin{figure}
    \centering
    \includegraphics{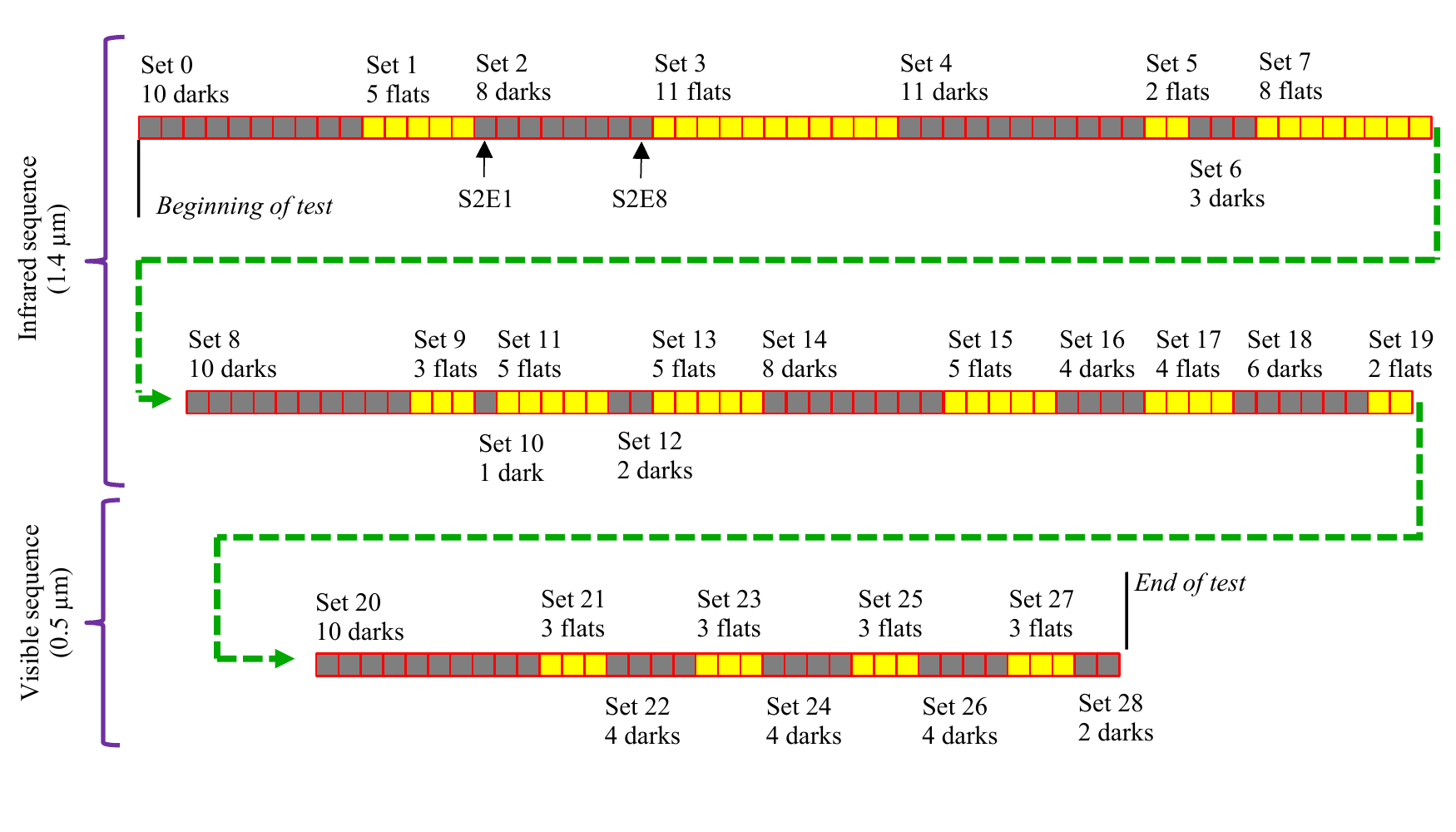}
    \caption{\label{fig:sequence-full}The sequence of flat and dark exposures used in this paper.}
\end{figure}

Each SCA goes through a sequence of flat and dark exposures during acceptance testing that is intended mainly for correlation function analyses and investigation of time-dependent nonlinear effects such as burn-in \citep{2020JATIS...6d6001M}. The sequence is shown in Fig.~\ref{fig:sequence-full}. The first part of the test is carried out with illumination at a wavelength of 1.4 $\mu$m, chosen since many of the {\slshape Roman} science programs make extensive use of $J$ and $H$ bands. The second part of the test (not used in our previous papers) is at a wavelength of 0.5 $\mu$m, where quantum yield effects are expected to be important. Exposures are identified with a ``Set'' number and an ``Exposure'' number within that set, e.g., S2E8 is the 8th exposure in Set 2. This second part contains 12 flat field exposures, reaching to near full well. Each exposure consists of 64 full frames at 2.75 seconds per frame.

These tests are carried out at an SCA temperature of 95 K and a bias voltage of 1.0 V. The SCA is read out in 32 channels, each consisting of 128 columns and 4096 rows. While the SCAs being tested are flight candidates, there were some differences relative to the flight setup: most notably, these tests were performed using a Leach controller instead of the flight electronics, and the guide window was not used (it will be used in flight).

\subsection{Quantum efficiency, full well, and noise data}
\label{ss:qe-data}

Another set of measurements is possible with the quantum efficiency (``{\tt qe}''), full well (``{\tt fw}''), and noise (``{\tt noise}'') data sets taken during acceptance testing. The quantum efficiency data consists of a set of flat fields at many wavelengths, ranging from 480 nm (the shortest science wavelength for the {\slshape Roman} wide field instrument) to beyond the $\sim 2.5$ $\mu$m cutoff of the detectors. This should enable determination of the 2-hole probability as a function of wavelength. However, the quantum efficiency data typically has only 2 flat fields per wavelength and 5 frames per data cube. This means that much more averaging needs to be done with this data than the flat + dark data optimized for correlation function analysis (\S\ref{ss:fd}). In order to ensure that the results are statistically independent from the flat + dark data, we pair the visible quantum efficiency data with infrared flat fields from the {\tt fw} data set (at a wavelength of 1.4 $\mu$m) and dark frames from the {\tt noise} data.

\subsection{Simulations}\label{simulations}

We also generate a simulated suite of data cubes, in which the charge diffusion parameters are known. These can be generated using a modification of the simulation code from \citet{2020PASP..132a4501H}, which we summarize here. The process begins by reading a user-provided configuration file containing simulation inputs and specifications. Using default settings generates a data cube of size $4096^2$ pixels$^{2}$ containing 66 time samples (each having 2 substeps) with the bounding 4 rows and columns designated as reference pixels.

Next, arrays are initialized and charge is drawn from a Poisson distribution and accumulated at each time step. When BFE is turned on, the Poisson distribution is modified at each time step based on pixel area defects. This does not happen in the absence of BFE; each time frame will have a random realization of charge accumulation. Once charge is accumulated over all time frames, it may be optionally convolved with an IPC kernel. If any CNL is desired, it can be applied following the inclusion of IPC. We add a small amount of Gaussian noise to our simulated data. In the final step, we convert charge to data numbers and save the output datacube as a \textsc{fits} file. We direct readers to Figure 1 of \citet{2020PASP..132a4501H} to see a flowchart of the simulation procedure.

The principal modification needed to the above process is that in a time step $\delta t$, we need to generate both 1-electron and 2-electron events. Let us first consider what happens without the BFE; in this case, we need to generate some 1-electron events, with a Poisson distribution in each pixel with mean $(1-\omega) I_\gamma\delta t$, and some 2-electron events, with a Poisson distribution in each pixel with mean $\omega I_\gamma\delta t$. The 1-electron events are straightforward. For the 2-electron events, we write a {\tt for} loop to run over the possible offsets $(\Delta i,\Delta j)$, and generate 2-electron events with Poisson distribution in each pixel with mean $\omega p_2(\Delta i,\Delta j)I_\gamma\delta t$. We can add both this array and the same array offset by $(\Delta i,\Delta j)$ to the map of total charge $Q$.

A complication occurs when we include the BFE, since then the pixels have areas that are not identical. Instead of shifting the pixel boundaries in the simulations and following each 2-electron event, we have instead applied the BFE only to the 1-electron events, and increased the magnitude of the pixel area correction by a factor of $1/f$, where $f=(1-\omega)/(1+\omega)$ (since the fraction of electrons coming from 1-electron events is $f$). This is not perfectly realistic, but it captures the correct effective change in area of the pixel, which is the mechanism by which the BFE will appear in the correlation functions.

To ensure realism in our simulations, we base simulation inputs and specifications on real data (or reports) provided by the Detector Characterization Lab (DCL) at NASA's Goddard Space Flight Center. This includes quantities such as time steps between sampled frames and quantum efficiency. After passing the data through our analysis code, {\sc Solid-waffle}, we can measure gain, illumination, the linear IPC coefficient, and BFE coefficients. We discuss our simulation inputs and outputs in \S\ref{sec:sim_tests}.

\section{Analysis code}\label{sec:analysis}

We have expanded the {\sc Solid-waffle} code to include quantum yield effects. First, the code performs its characterization using the IR flats (+darks) to get the gain $g$, classical non-linearity coefficients $\beta_j$, and IPC coefficients $\alpha_{\rm H}$, $\alpha_{\rm V}$, and $\alpha_{\rm D}$, using the same procedure as in \citet{2020PASP..132g4504F}. These detector effects operate on already collected charge and should not depend on the wavelength of illumination. We then turn our attention to the visible flats (+darks), which we can use to determine the BFE kernel $[K^2a]$ (which could depend on absorption depth and hence wavelength since it acts during the charge collection process), as well as $\omega$ and $p_2$ and the current $I_e$ used in the visible flats ($I_e$ is not a main output of the characterization process, but must be solved in an auxiliary sense).

In what follows, we describe the sequence of steps for the visible flats. The steps are done in each super-pixel, regions of the detector used for analysis over which we measure a correlation function. In most analyses, there are 1024 super-pixels, each composed of $128\times 128$ pixels, but {\sc Solid-waffle} allows for any power of 2 and for different binnings in the row and column directions if desired.

\subsection{Determining $I_e$}

For each super-pixel, we construct the synthetic correlated double sample images (see \citealt{2020PASP..132a4501H} for a discussion on their importance) $S_{ad}(x_1,x_2|{\rm F}k)$ for each flat $k$, where $a$ and $d$ are the first and last time slices used for visible characterization. We then take the super-pixel median and subtract the median of the reference pixels for that flat exposure. The default scheme, which we have found to work well on data from the HyC dewar at the DCL, is to use the pixels on the left and right sides; for $128\times 128$ super-pixels, this corresponds to the median of 1024 reference pixels, 512 on each side. We take the mean over the flats to get an averaged signal $\bar S_{ad}$ for that super-pixel. This is related to the photocurrent $I_e$ by
\begin{equation}
\bar S_{ad} = \frac1g\Bigl\{ I_et_{ad} - \sum_{j=2}^p \beta_j [(I_et_d)^j - (I_et_a)^j] \Bigr\},
\end{equation}
or, with some algebraic rearrangement,
\begin{equation}
I_e = \frac{ g\bar S_{ad} }{ t_{ad} - \sum_{j=2}^p \beta_j I_e^{j-1} (t_d^j-t_a^j) }.
\end{equation}
We use this equation iteratively to solve for $I_e$ and use a starting guess of $I_e = g\bar S_{ad}/t_{ad}$ (i.e., without the nonlinearity).

We have experimented with other ways to estimate the current $I_e$. The aforementioned method is more robust than methods that used, e.g., the linear coefficient of a polynomial fit, since the latter can suffer from persistence or other effects that are most important in the first couple of frames.

\subsection{Measuring the correlation function}

We next want to measure a correlation function that contains the information in $\Phi$. If we want to constrain a $5\times 5$ even\footnote{$\Phi$ is even in the sense of $\Phi(-\Delta x_1,-\Delta x_2) = \Phi(\Delta x_1,\Delta x_2)$.} kernel $\Phi$, then we need a measurement that is also a $5\times 5$ even kernel ${\boldsymbol v}$. We choose the same correlation functions that were used in the ``Advanced'' characterization in \citet{2020PASP..132a4501H}. That is, we build a $5\times 5$  array ${\boldsymbol v}$ of the averaged equal-interval correlation functions of duration $\mu$, built from time slices between $a$ and $d$:
\begin{equation}
{\boldsymbol v}(\Delta x_1,\Delta x_2) = \bar C_{a,a+\mu,a,a+\mu,[d-a-\mu]}(\Delta x_1,\Delta x_2)
= \frac1{d-a-\mu}\sum_{k=0}^{d-a-\mu-1} C_{a+k,a+\mu+k,a+k,a+\mu+k}(\Delta x_1,\Delta x_2)
\end{equation}
for $(\Delta x_1,\Delta x_2) \neq (0,0)$. The equal-interval correlation function should be sensitive to the probability of generating pairs of charges in pixels $(x_1,x_2)$ and $(x_1+\Delta x_1,x_2+\Delta x_2)$. The averaging reduces noise. For the special case of the correlation function at $(0,0)$, we want to remove the read noise; we thus set
\begin{equation}
{\boldsymbol v}(0,0) = \Delta V = \bar C_{a,a+\mu,a,a+\mu,[d-a-\mu]}(0,0)
- \bar C_{a,a+\mu',a,a+\mu',[d-a-\mu']}(0,0),
\label{eq:V00}
\end{equation}
where $\mu'<\mu$ (typically we choose $\mu'=1$, $\mu=3$).\footnote{In the {\sc Solid-waffle} visible flat portion of the code, $a$ is represented by {\tt ts\_vis-basicpar.reset\_frame}; $d$ by {\tt te\_vis-basicpar.reset\_frame}; $d-a-\mu$ by {\tt nvis}; $\mu'$ by {\tt tchar1\_vis}; $\mu$ by {\tt tchar2\_vis}; and ${\boldsymbol v}$ by {\tt corr\_mean}.}

\subsection{Iteratively solving for quantum yield, charge diffusion, and IPNL parameters}\label{sec:iterative}

We now enter an iterative loop where $\Phi$ and the IPNL kernel $[K^2a]$ are determined. In each iteration, we first update $\Phi$ and then $[K^2a]$.

The update to the vector ${\boldsymbol\Phi}$ uses an approximate Newton-Raphson iterative step:
\begin{equation}
    {\boldsymbol\Phi}_{n+1} = {\boldsymbol\Phi}_n + {\bf J}^{-1}_{\textrm{approx}} \left[ {\boldsymbol v}({\rm measured}) - {\boldsymbol v}({\rm predicted~from~}{\boldsymbol\Phi}_n) \right].
\end{equation}
where ${\boldsymbol\Phi}_n$ is the ``vector'' representing the kernel $\Phi$ in the $n$th iteration, and ${\bf J}_{\textrm{approx}}$ is an approximation to the Jacobian $\partial {\bm v}/\partial \Phi$. In this equation, ${\boldsymbol\Phi}$ and ${\boldsymbol v}$ are understood to be ``flattened'' length-25 vectors, and the Jacobian is a $25\times 25$ matrix. The predicted ${\boldsymbol v}$ is obtained from {\tt ftsolve.solve\_corr\_vis\_many} (two calls are necessary because of the second term in Eq.~\ref{eq:V00}). The Newton-Raphson method would use the exact Jacobian, but the method converges even with an approximate Jacobian (this saves the time associated with evaluating the Jacobian, but at the expense of requiring more iterations to converge).

We construct our approximate Jacobian as follows. We determine the flat field power spectrum for quantum yield and charge diffusion only (no BFE, IPC, or CNL) by taking the BFE-only case (Eq.~\ref{eq:unequal-time}) and considering the limit when BFE goes to zero:
\begin{equation}\label{eq:noBFE_FT}
    C^{\Phi~\rm only}_{abcd}(k_1'-k_1,k_2'-k_2) =
    \lim_{\textrm{BFE} \to 0} \widetilde{C}^\mathrm{BFE}_{abcd}(k_1'-k_1,k_2'-k_2)= \frac{I_et_{abcd}N^2\delta_{k+k',0}}{g^2}\left[2\tilde{\Phi}(k_1',k_2')+1\right].
\end{equation}
(We used l'H\^opital's rule, and used the overlap $t_{abcd}$ as defined in \citet{2020PASP..132a4501H}[Eq.~18].) Data from flats and darks are taken and analyzed in real space, so we Fourier transform Eq.~(\ref{eq:noBFE_FT}), use the fact that $\Phi$ is even, and then find the correlation function
\begin{equation}
    C^{\Phi~\rm only}_{abcd}(x_1-x'_1,x_2-x'_2)=\frac{I_e t_{abcd}}{g^2}\left[2\Phi(x_1-x'_1,x_2-x'_2)+\delta_{x_1,x_1'}\delta_{x_2,x_2'}\right].
\end{equation}
This suggests that we use the approximate Jacobian
\begin{equation}
J_{\rm approx}(\Delta x_1,\Delta x_2; \Delta y_1, \Delta y_2)
= \frac{\partial v^{\Phi~\rm only}(\Delta x_1,\Delta x_2)}{\partial \Phi(\Delta y_1,\Delta y_2)}
= \frac{2I_e t_{abcd}}{g^2} \delta_{\Delta x_1,\Delta y_1} \delta_{\Delta x_2,\Delta y_2}
\times \left\{ \begin{array}{lll}
\mu & & (\Delta x_1,\Delta x_2)\neq (0,0) \\
\mu-\mu' & & (\Delta x_1,\Delta x_2)=(0,0)
\end{array}\right..
\end{equation}
Because ${\bf J}_{\rm approx}$ is diagonal, the inverse is trivial it can be implemented with element-wise multiplication, rather than explicitly flattening ${\boldsymbol\Phi}$ and ${\boldsymbol v}$ into vectors, multiplying by a matrix, and turning them back into {\tt numpy} arrays.

The update of the IPNL kernel $[K^2a]$ (here assumed to be mainly BFE) is performed using the same algorithm as in \citet[\S4.2.2]{2020PASP..132g4504F}. The only change necessary to the function that implements this algorithm ({\tt pyirc.bfe}) is to call the new visible correlation function {\tt ftsolve.solve\_corr\_vis} instead of {\tt ftsolve.solve\_corr}, and to put in the piping to ensure that {\tt ftsolve.solve\_corr\_vis} has access to the current values of $\omega$ and $p_2$.

Once $\Phi(\Delta x_1,\Delta x_2)$ is measured, we may solve for the values of $(\omega,C_{11},C_{12},C_{22})$ that minimize the sum of squared residuals assuming a Gaussian charge diffusion model (\S\ref{sec:Gauss}). In the presence of noise, we impose a constraint that ${\bf C}$ be positive definite; in practice, we restrict the minimum eigenvalue of ${\bf C}$ to be $\ge (0.01\,{\rm pix})^2$, i.e., the semiminor axis of the charge diffusion must be at least 0.01 pix.

\subsection{Tests on simulations}\label{sec:sim_tests}
\begin{figure}
    \centering
    \includegraphics[width=\textwidth]{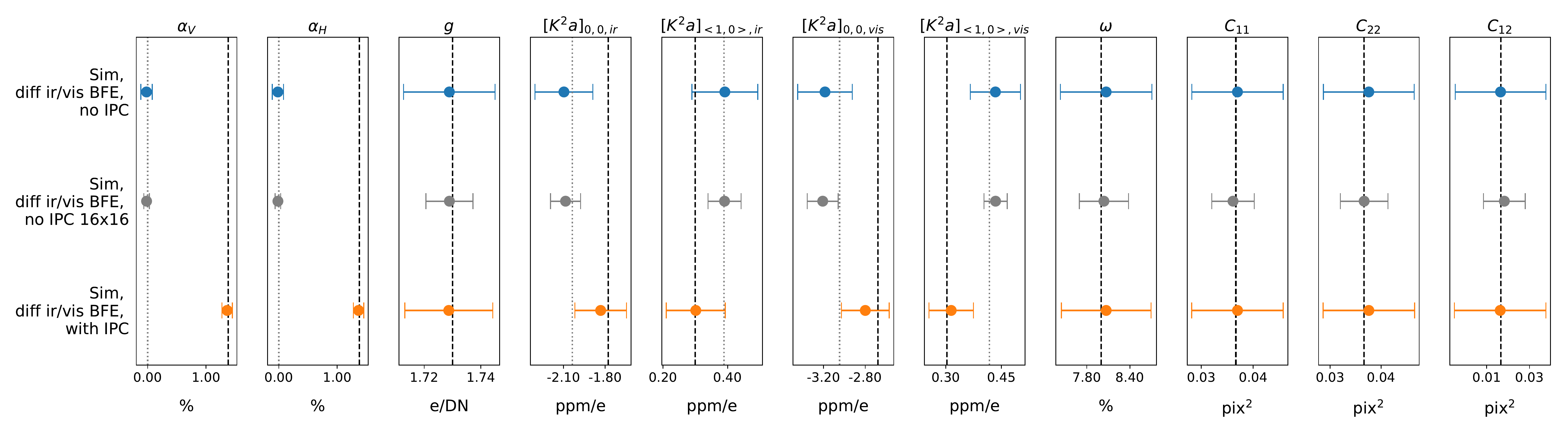}
    \caption{\label{fig:paramplot-sim} Simulation tests. Each row corresponds to a simulation variation. The vertical lines indicate input simulation values. For the panels that have two dashed vertical lines, the long-dashed line indicates the input values corresponding to simulation variations with non-zero IPC.}
\end{figure}

As was done in \citet{2020PASP..132a4501H} and \citet{2020PASP..132g4504F}, we created a set of simulated flats and darks (both visible and IR) following the procedure outlined in \S\ref{simulations}. All results in this section are from data given in $32 \times 32$ bins of superpixels. Figure~\ref{fig:paramplot-sim} summarizes the simulation tests. The first row does an analysis for 32$\times$32 superpixels, assumes no IPC, sets input BFE kernels to be different for the IR and visible data (with a stronger impact in the visible data), and a fiducial set of quantum yield and charge diffusion parameters. The second row is similar to the first but with 16$\times$16 superpixels. The third row is similar to the first but with an input IPC. For all three tests, the input gain was $g=1.7285$ e/DN and {\sc Solid-waffle} returned a consistent output, with the third test giving $g=1.7287 \pm 0.0004$ e/DN. We fit the classical nonlinearity curve up to quartic order using polynomial coefficients $\beta_j$ (units: electrons$^{1-j}$) defined in F/37. We set $\beta_j=\{0\}$ and {\sc Solid-waffle} recovered values of $g\beta_2,\, g^2\beta_3,\textrm{ and } g^3\beta_4$ consistent with zero at $1\sigma$. The charge per time slice we passed to the simulation is $3051.5$ e and the output was $2898.9 \pm 0.8$ e; this result is consistent with the input quantum efficiency of 95\%. We passed IPC components $\alpha_H = \alpha_V = \alpha = 0.01379$ and obtained $\alpha_H = 0.01367 \pm 0.00003,\, \alpha_V = 0.01363 \pm 0.00003$. These correspond to biases of $\sim 1\%$ compared to the inputs. Results discussed thus far are broadly in agreement with those in \citet{2020PASP..132g4504F} which is promising (but unsurprising) since CNL and IPC are unaffected by quantum yield.

For the remainder of this section we review results that depend on quantum yield and charge diffusion parameters. The value of $\omega$ input to {\sc Solid-waffle} was $0.08$ and it returned $\omega=0.0807 \pm 0.0002$. Input values for the charge diffusion covariance matrix were $C_{11}=C_{22}=0.04 \textrm{ and } C_{12}=0.02$. Output values were $C_{11}=0.0405 \pm 0.0004,\, C_{11}=0.0414 \pm 0.0005, \textrm{ and } C_{12}=0.0197 \pm 0.0004$.  

We now discuss the differences between {\sc Solid-waffle} simulation inputs and outputs for IPNL. The central pixel IR IPNL kernel returned by {\sc Solid-waffle} is $[K^2a']_{0,0}=-1.8321 \pm 0.0058$ (stat) ppm/e, which we compare to the input value $[K^2a']_{0,0,\textrm{input}}=-1.7779$ ppm/e. This is a bias of $-0.0542$ ppm/e ($3\%,\,9.3\sigma$) compared to the input value. {\sc Solid-waffle} gives comparable results for visible simulations. The recovered visible IPNL kernel is $[K^2a']_{0,0}=-2.7995 \pm 0.0072$ (stat) ppm/e while the input value is $[K^2a']_{0,0,\textrm{input}}=-2.6777$ ppm/e. This corresponds to a difference of $-0.1218$ ppm/e ($4.5\%,\,17\sigma$). We also tested our code's ability to recover the IPNL kernels of symmetrically-averaged nearest neighbor pairs of pixels. We found $[K^2a']_{\langle 1,0\rangle }=0.3016 \pm 0.0055$ (stat) ppm/e for the IR IPNL kernel. Compared to its input value $[K^2a']_{\langle 1,0\rangle ,{\rm input}}=0.3001$ ppm/e, this represents a bias of $0.0015$ ppm/e ($0.50\%,\,0.28\sigma$). Similarly, in the visible case we recovered $[K^2a']_{\langle 1,0\rangle }=0.3148 \pm 0.0036$ (stat) ppm/e which can be compared to the given value $[K^2a']_{\langle 1,0\rangle ,{\rm input}}=0.3033$ ppm/e. This translates to a bias of $0.0115$ ppm/e ($3.8\%,\,3.2\sigma$). We do not find significant differences in results when using $16 \times 16$ bins of superpixels.

For the IR case we can directly compare the above IPNL kernel results to those in \citet{2020PASP..132g4504F}. Therein, the authors found a $0.96\%$ bias between inputs and outputs for the central pixel and a bias of $1.5\%$ for the symmetrically-averaged nearest neighbors. Those percentages are similar to what we find in this work, 3\% for central pixel and 0.5\% for nearest neighbors.

\section{Results for {\slshape Roman} detector arrays}\label{sec:results}
We now apply our updated formalism to {\slshape Roman} flight candidates SCAs 20663, 20828, and 20829. Details of the fiducial set of visible detector runs we use in this paper are given in Tables \ref{frames_20663} -- \ref{tab:config_vis_fid}. Following \citet{2020PASP..132g4504F}, some results will only be provided for SCA 20829 for brevity. In the future we expect to set up a public repository where other results will be made available. 

\subsection{Superpixel maps for SCA 20829}

\begin{figure}
    \centering
    \includegraphics[width=0.8\textwidth]{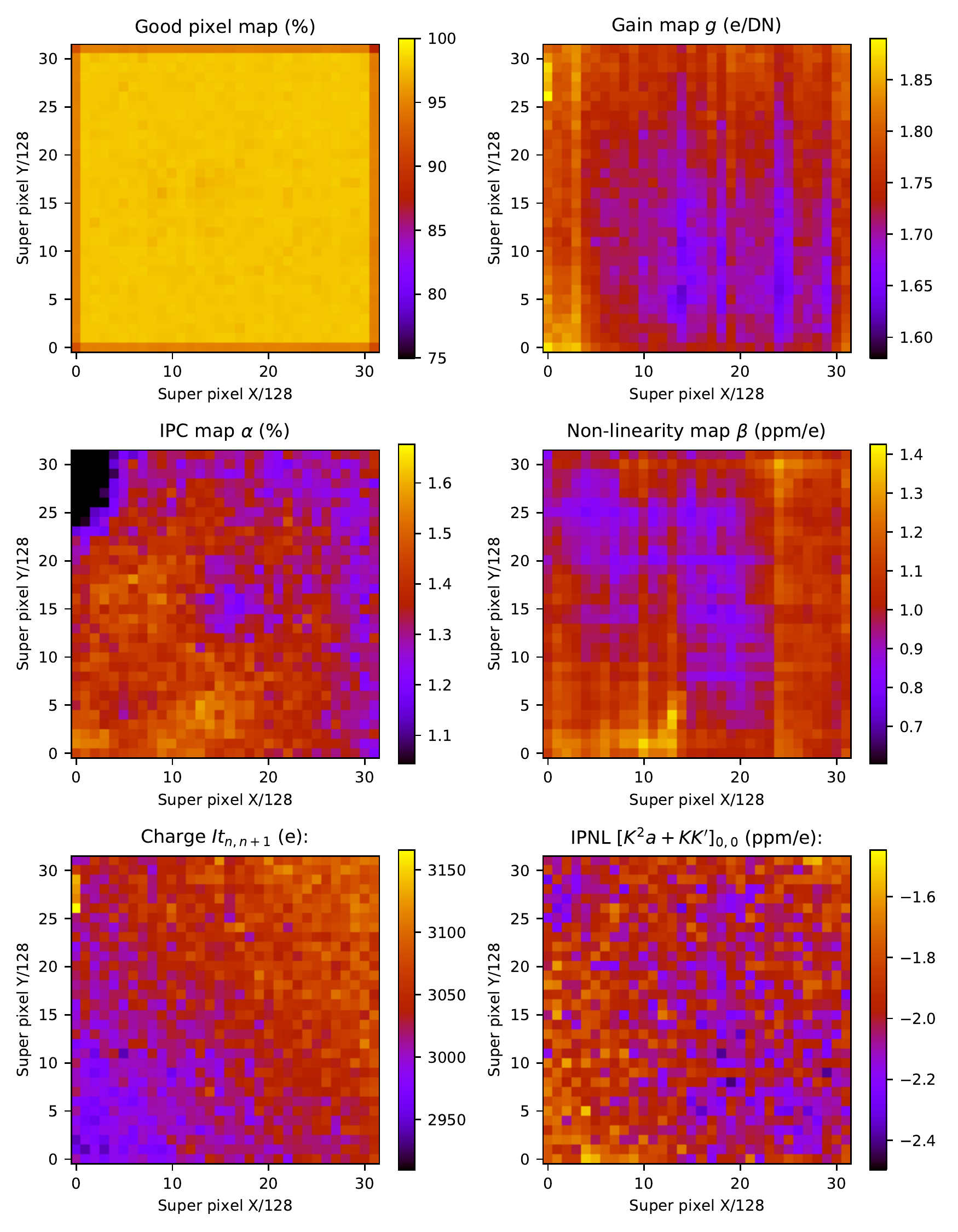}
    \caption{\label{fig:pixel_maps}Detector maps showing the spatial variation of IR characterization parameters for SCA 20829. This is binned into 1024 superpixels each composed of $128 \times 128$ pixels. These maps agree with their counterparts (when present) in \citet{2020PASP..132g4504F}. $\beta$ shown in the center right map is $\beta_2$ from Equation \ref{eq:Sa}. The nonlinearity polynomial coefficients $\beta_j$ are strongly anticorrelated (correlation coefficient $-0.94$ between $\beta_2$ and $\beta_3$, and $-0.96$ between $\beta_3$ and $\beta_4$), so large variations seen in $\beta_2$ correspond to smaller variations in the total nonlinearity.}
\end{figure}

\begin{figure}
    \centering
    \includegraphics[width=\textwidth]{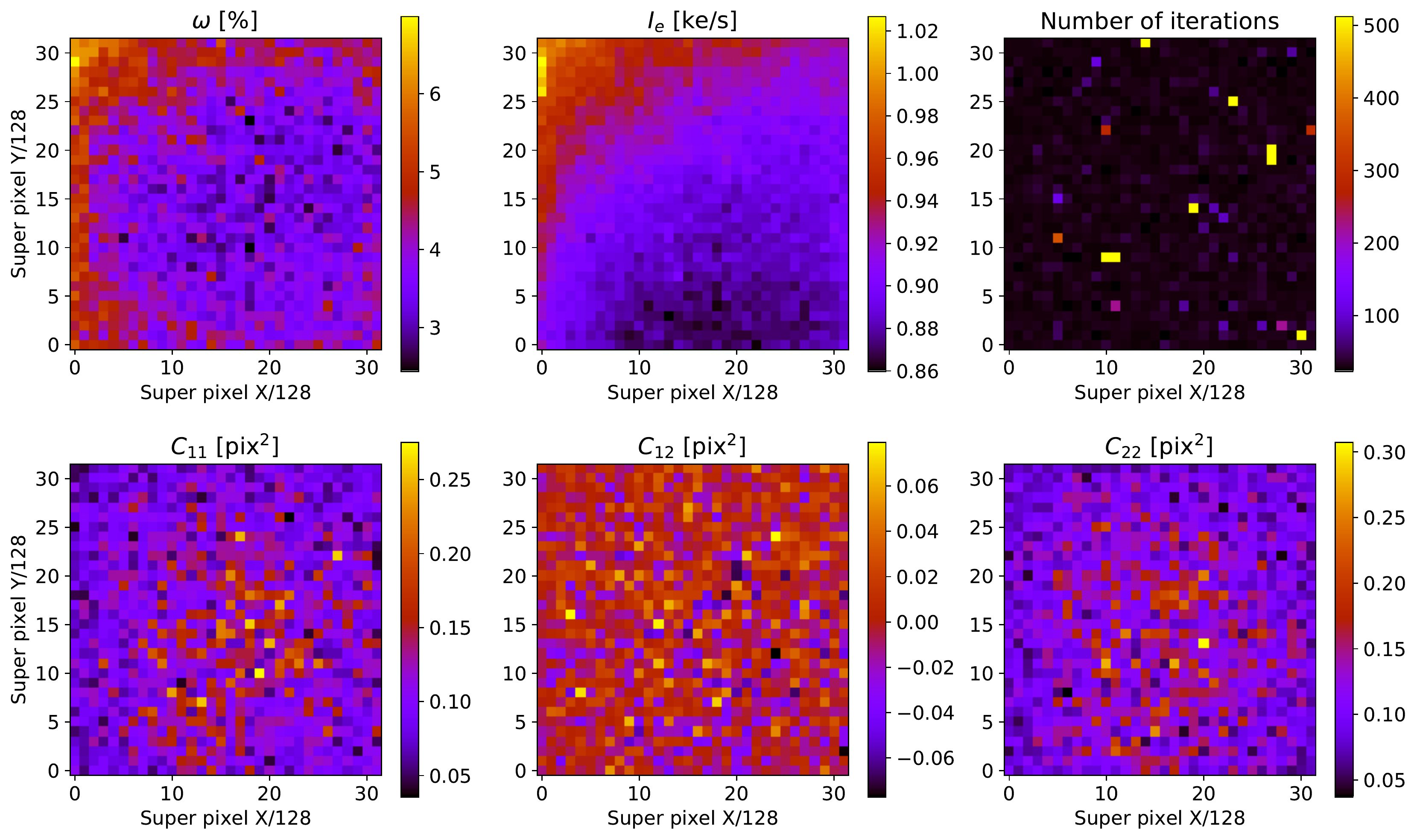}
    \caption{\label{fig:pixel_maps_cd} Detector maps showing the spatial variation of visible characterization parameters for SCA 20829. This is binned into 1024 superpixels each composed of $128 \times 128$ pixels. Number of iterations refers to the number of steps required for the $\Phi$ kernel to converge (see \S\ref{sec:iterative}).}
\end{figure}

Figure \ref{fig:pixel_maps} shows the spatial distributions of derived IR characterization quantities across SCA 20829, and Figure \ref{fig:pixel_maps_cd} does the same for derived visible quantities. Figure \ref{fig:pixel_maps} is very similar to Figure 8 of \citet{2020PASP..132g4504F}, which is expected because there are no changes to this part of the measurement and relatively minor changes to the set-up (time frames used for the correlation analysis and non-linearity curve fitting and different input flats and darks). A few important points should be noted about these data. First, the illumination pattern, while smooth, is not exactly flat, so the gradient in charge per frame ($It_{n,n+1}$) is not necessarily a property of the SCA. Second, the probability $\omega$ of generating two electron-hole pairs is highest in the upper left corner of the detector. We have not identified the cause of this behavior, and it is not seen in the other SCAs. Last, the $C_{11}$ and $C_{22}$ components of the charge diffusion covariance matrix are the same to within 10\%. This is convenient for weak lensing applications, since $C_{11}-C_{22}$ is a source of ellipticity ($e_1$ component). It is expected behavior since the electric field structure in the HgCdTe layer should not distinguish the $x$ vs.\ $y$ axes. This can be contrasted with CCDs, where there is a readout direction and there are generally higher potential barriers between adjacent columns than between adjacent rows.

\subsection{Recovering detector parameters}
\label{ss:recover}

\begin{figure}
    \centering
    \includegraphics[width=0.8\textwidth]{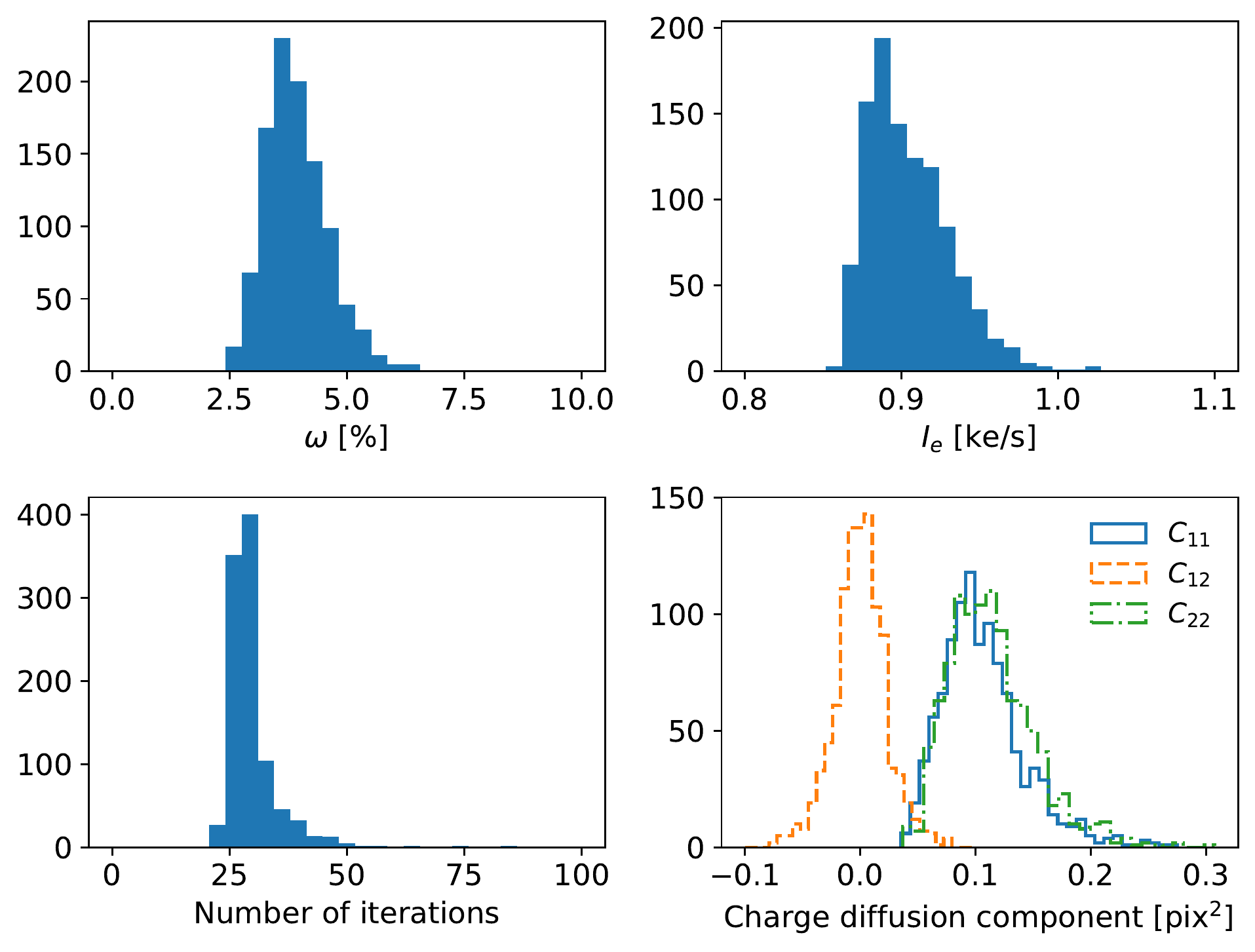}
    \caption{\label{fig:cd_quantities}Histograms of visible characterization results for SCA 20829. The $y$-axis counts the number of superpixels falling into each bin. Number of iterations refers to the number of steps required for the $\Phi$ kernel to converge.}
\end{figure}

\begin{figure}
    \centering
    \includegraphics[width=1.0\textwidth]{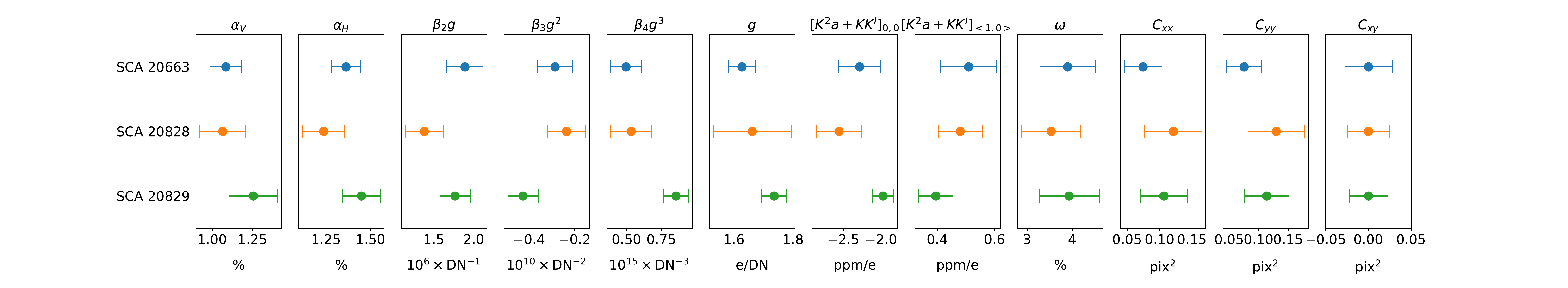}
    \caption{\label{fig:paramplot_vis}Fiducial characterization  results  for each SCA  averaged  over  all  superpixels (see Table \ref{tab:char_outputs_all} for the numerical values for each SCA). Note that the IPNL values shown are for the visible case, although the `vis' tag has been omitted here for readability.}
\end{figure}

Mean values and standard deviations on the mean (both taken across all superpixels) for infrared IPNL, visible IPNL, and IPC parameters are given in Table \ref{tab:char_outputs_all} for the three flight candidates. Figure \ref{fig:cd_quantities} offers a visual representation of quantum yield and charge diffusion properties for SCA 20829, while Figure \ref{fig:paramplot_vis} summarizes most data from Table~\ref{tab:char_outputs_all} as a whisker plot. 

First, we compare IR characterization results from our analyses to those reported in the fidicual results of \citet{2020PASP..132g4504F}\footnote{Our convention is to use standard deviations calculated in this work when comparing mean values from these two investigations.}. Values of charge per time slice, gain, and all four IPC $\alpha$ parameters are consistent at the $<1\sigma$ level for all detectors. The large-scale variation in gain across SCA 20829, as shown in the upper right of Figure \ref{fig:pixel_maps}, is consistent between the two works, and similar variations are seen in other {\slshape Roman} flight candidates at the DCL. The CNL parameters $\beta_jg^{1-j}$ between the two works are significantly different for all detectors, with differences ranging from $4.0-9.3\sigma$ for 20663, $3.3-8.2\sigma$ for 20828, and $5.6-16.2\sigma$ for 20829, and increasing as $j$ increases.  However, the nonlinearity curves were fit over different time ranges for the two papers, which could understandably lead to the large differences in the fitted $\beta_j$. Thus, it may be more instructive to compare the mapping between charge and signal following F/32 which gives 
\begin{equation}\label{eq:Sa}
    S_a = \frac{1}{g}\left[\bar Q(t_a)-\beta_2 \bar Q(t_a)^2-\beta_3 \bar Q(t_a)^3-\beta_4 Q(t_a)^4\right].
\end{equation}
At time frame 10, $S_{10}=18765.19 \pm 17.13$ \citep{2020PASP..132g4504F} and $S_{10}=18754.38 \pm 23.63$ (this work), where errors are obtained by propagating the standard errors from the charge per time slice, gain, and nonlinearity parameters. These signal values are entirely consistent. Finally, neighboring pixel IPNL kernel results for each detector are consistent with those reported in \citet{2020PASP..132g4504F} at $\lesssim 1\sigma$. However, the central pixel results are discrepant at $\sim 2\sigma$ in each case. The discrepancy may be a result of this work and \citet{2020PASP..132g4504F} using different sequences of flats and darks as well as different time slices.

We now compare visible characterization results between detectors and compare IR and visible IPNL kernels for each detector. SCA 20663 has the greatest current per pixel of the flight candidates, beating out SCA 20828 by a factor of 1.45 and SCA 20829 by a factor of 1.53. We suspect that this is due to differences in light source intensity hitting each detector. All three detectors show evidence of $\omega > 0$ at the $5-6\sigma$ level for each superpixel. For every photon absorbed by these detectors, more than 1.035 charge carriers are collected on average. The charge diffusion variance contributions $C_{11}$ and $C_{22}$ are detected at $3\sigma$ for each SCA while the covariance term ($C_{12}$, describing whether there is more diffusion in the northeast-southwest direction than the northwest-southeast direction on the SCA) is consistent with zero. We have also shown the charge diffusion length, which we have given as the $1\sigma$ scale of a 2-dimensional Gaussian:
\begin{equation}
\sigma_{\rm cd} = (10\,\mu{\rm m})\times\sqrt{\frac{C_{11}+C_{22}}2}.
\end{equation}
Here the factor of 10 $\mu$m converts from pixel units to physical length. The array average charge diffusion lengths reported are 2.7, 3.5, and 3.3 $\mu$m for SCAs 20663, 20828, and 20829, respectively.

In all three detectors we see that the visible central pixel IPNL kernel has a larger absolute value than its IR counterpart, but this difference is consistent with zero after considering the standard deviation. The rate of decline in the absolute value of the IPNL kernel as one goes from $(0,0)$ to $\langle 1,0 \rangle$ to $\langle 2,0\rangle$ is approximately the same for IR and visible characterizations.

\begin{table}
\scriptsize  
    \centering
    \begin{tabular}{llrrrrrr}
    \hline \hline
    & & \multicolumn{2}{c}{SCA 20663}
    & \multicolumn{2}{c}{SCA 20828}
    & \multicolumn{2}{c}{SCA 20829}\\
    \hline
    Quantity     &Units &fid &$\sigma({\rm fid})$ &fid &$\sigma({\rm fid})$ &fid &$\sigma({\rm fid})$  \\
    \hline
   
    \hline
\multicolumn{8}{c}{IR characterization} \\    
    \hline
    Charge, $It_{n,n+1}$     &ke                        &3.4918 &0.0433  & 3.0816 &0.0319  & 3.0376 &0.0333  \\
    Gain $g$                 &e/DN                      &1.6264 &0.0444   & 1.6696 &0.0618  & 1.7357 &0.0420   \\
    IPC $\alpha$             &\%                        &1.2234 &0.0743  & 1.1508 &0.1085  & 1.3520 &0.1077  \\
    IPC $\alpha_{\rm H}$     &\%                        &1.3624 &0.0808   & 1.2363 &0.1185   & 1.4483 &0.1068   \\
    IPC $\alpha_{\rm V}$     &\%                        &1.0844 &0.0991  & 1.0653 &0.1420  & 1.2556 &0.1508  \\
    IPC $\alpha_{\rm D}$     &\%                        &0.1231 &0.0478  & 0.1215 &0.0463  & 0.1407 &0.0423  \\
    $\beta_2g$               &$10^6\times$DN$^{-1}$     &1.8884 &0.2270   & 1.3793 &0.2391   & 1.7637 &0.1889   \\
    $\beta_3g^2$             &$10^{10}\times$DN$^{-2}$  &$-$0.2857 &0.0780   & $-$0.2354 &0.0840   & $-$0.4255 &0.0662   \\
    $\beta_4g^3$             &$10^{15}\times$DN$^{-3}$  &0.4966 &0.1109  & 0.5329 &0.1471  & 0.8553 &0.0895  \\
    
    $[K^{2}a + KK^I]_{0,0\textrm{, IR}}$   &ppm/e     &$-$2.2839 & 0.2823   & $-$2.5582 & 0.3050   & $-$1.9714 & 0.1422   \\
    $[K^{2}a + KK^I]_{\langle 1,0\rangle \textrm{, IR} }$ &ppm/e     &0.5091 &0.0975  & 0.4800 &0.0769  & 0.3946 &0.0596  \\
    $[K^{2}a + KK^I]_{\langle 1,1\rangle \textrm{, IR}}$ &ppm/e     &0.1626 &0.0871  & 0.1351 &0.0676  & 0.1117 &0.0610  \\
    $[K^{2}a + KK^I]_{\langle 2,0\rangle \textrm{, IR}}$ &ppm/e     &0.0561 &0.0813   & 0.0219 &0.0607   & 0.0303 &0.0608   \\
    $[K^{2}a + KK^I]_{\langle 2,1\rangle \textrm{, IR}}$ &ppm/e     &0.0358 &0.0651  & 0.0098 &0.0445  & 0.0151 &0.0447  \\
    $[K^{2}a + KK^I]_{\langle 2,2\rangle \textrm{, IR}}$ &ppm/e     &0.0254 &0.0701  & 0.0022 &0.0657  & 0.0055 &0.0632  \\
    $[K^{2}a+KK^I]_{\rm H \textrm{, IR}}$   &ppm/e     &0.4996 & 0.1182   & 0.4644 & 0.1003   & 0.3742 & 0.0858   \\
    $[K^{2}a+KK^I]_{\rm V \textrm{, IR}}$   &ppm/e     &0.5186 & 0.1072  & 0.4957 & 0.0995  & 0.4150 & 0.0813  \\ \hline
    \multicolumn{8}{c}{Visible characterization} \\    \hline

    $I_e$                   &ke/s                     &1.3821 &0.0286   & 0.9522 &0.0251   & 0.9048 &0.0264   \\
    $\omega$                &\%                       &3.8939 &0.6118  & 3.5296 &0.6568  & 3.9290 &0.6646  \\
    $C_{11}$                &pix$^2$                 &0.0745 &0.0293   & 0.1212 &0.0441   & 0.1066 &0.0364   \\
    $C_{12}$                &pix$^2$                 &0.0001 &0.0275  & 0.0000 &0.0245  & 0.0001 &0.0227  \\
    $C_{22}$                &pix$^2$                 &0.0754 &0.0293  & 0.1299 &0.0479  & 0.1136 &0.0376  \\
    $\sigma_{\rm cd}$ & $\mu$m & 2.7061 & 0.4121 & 3.5180 & 0.4878 & 3.2890 & 0.4341 \\
    $[K^{2}a + KK^I]_{0,0\textrm{, vis}}$   &ppm/e     &$-$2.6909 & 0.5960   & $-$2.6048 & 0.3426   & $-$2.0872 & 0.1780   \\
    $[K^{2}a + KK^I]_{\langle 1,0\rangle \textrm{, vis} }$ &ppm/e     &0.5844 & 0.0860  & 0.4713 & 0.0953  & 0.3765 & 0.0808  \\
    $[K^{2}a + KK^I]_{\langle 1,1\rangle \textrm{, vis}}$ &ppm/e     &0.1517 &0.0757  & 0.1312 &0.0903  & 0.1044 &0.0847  \\
    $[K^{2}a + KK^I]_{\langle 2,0\rangle \textrm{, vis}}$ &ppm/e     &0.0252 &0.0713   & 0.0221 &0.0875   & 0.0188 &0.0830   \\
    $[K^{2}a + KK^I]_{\langle 2,1\rangle \textrm{, vis}}$ &ppm/e     &0.0077 &0.0516  & 0.0131 &0.0599  & 0.0074 &0.0594  \\
    $[K^{2}a + KK^I]_{\langle 2,2\rangle \textrm{, vis}}$ &ppm/e     &0.0011 &0.0762  & -0.0028 &0.0874  & 0.0007 &0.0847  \\
    $[K^{2}a+KK^I]_{\rm H \textrm{, vis}}$   &ppm/e     &0.5669 &  0.1135   & 0.4509 & 0.1286   & 0.3521 & 0.1151   \\
    $[K^{2}a+KK^I]_{\rm V \textrm{, vis}}$   &ppm/e     &0.6018 & 0.1152  & 0.4918 & 0.1237  & 0.4009 & 0.1148  \\
    
    \hline
    \end{tabular}
    \caption{Characterization results for SCAs 20663, 20828, and 20829 averaged over all superpixels. Rows refer to the fiducial configuration as described in Table \ref{tab:config_vis_fid}. Subscripts H and V on the IPNL kernel indicate the average of the horizontal and vertical nearest neighbors to the central pixel. Subscripts in angled brackets indicate averages over nearest neighbors, averages over diagonal neighbors, etc. For each SCA, the first column lists the value of that quantity in the fiducial run, and the second column lists the standard deviation of the superpixels for each quantity. The top portion of this table comes from IR characterization while the bottom portion comes from visible characterization.}
    \label{tab:char_outputs_all}
\end{table}

\subsection{Quantum yield vs.\ wavelength}
\label{sec:qy vs wavelength}

The quantum efficiency dataset (``{\tt qe}'') can be used to measure the quantum yield as a function of wavelength. Several modifications to the configuration were needed to achieve this. First, since there are only 2 flats at each wavelength, we made larger super-pixels, each consisting of $128\times 512$ pixels (so that there are 256 super-pixels total, each one readout channel wide, and with 8 super-pixels along the vertical direction in each channel). Second, since there are only 5 time stamps and the signal level is sometimes low, the {\tt qe} data do not constrain the BFE kernel; we used the {\sc Solid-waffle} keyword {\tt COPYIRBFE}, which sets the visible kernel $[K^2a]$ equal to that obtained from the IR flats rather than fitting it from the correlation function of the visible data. Finally, the fitting of the covariance matrix and quantum yield parameter from the $\Phi$ kernel is unstable when $\omega$ is small and the signal:noise ratio is low; therefore, we have estimated the 2-hole probability from
\begin{equation}
\omega(\lambda) = \frac{\sum_{\Delta x_1,\Delta x_2} \Phi(\Delta x_1,\Delta x_2)}{1 - \sum_{\Delta x_1,\Delta x_2} \Phi(\Delta x_1,\Delta x_2)}.
\end{equation}
(this is Eq.~\ref{eq:Phi-def}, summed using the normalization $\sum p_2=1$, and after solving for $\omega$). This form is numerically stable even if $\omega$ fluctuates to a negative value. We report both the median of $\omega$ over all 256 super-pixels, and the standard error on the median.

\begin{figure}
\centering
\includegraphics[width=6.75in]{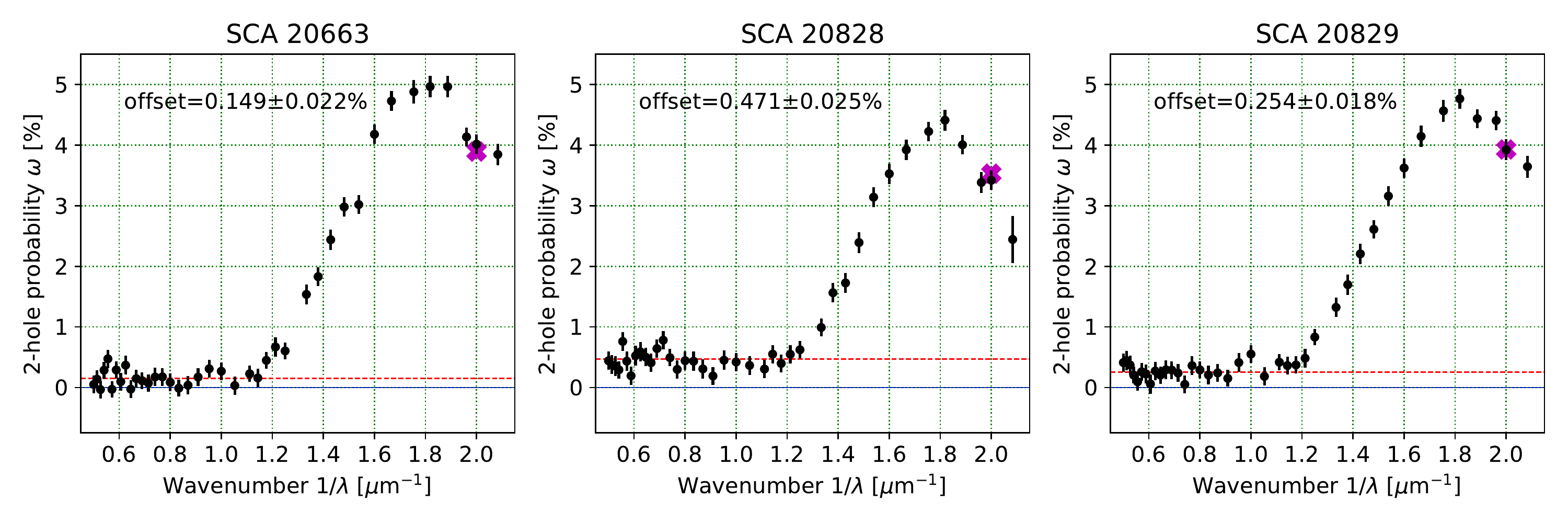}
\caption{\label{fig:omega_lambda}The dependence of the probability $\omega$ of producing 2 holes as a function of wavelength. The black points with error bars are the results from the {\tt qe} data set. The magenta X denotes the results from Table~\ref{tab:char_outputs_all} from the 500 nm flats ($\lambda^{-1} = 2.0\, \mu$m$^{-1}$). The probability peaks at $\lambda^{-1} \sim 1.8\,\mu$m$^{-1}$ and declines again toward bluer wavelengths. The probability $\omega$ should go to zero at redder wavelengths; we find an offset (reported in the caption) that is likely due to systematic errors (see main text).}
\end{figure}

Results are shown in Fig.~\ref{fig:omega_lambda} for all 3 SCAs. The probability of collecting 2 holes rises as a function of photon energy; we observe a threshold at $\lambda^{-1}\approx 1.2\,\mu$m$^{-1}$ ($\lambda \approx 830$ nm), where the photon has approximately 3 times the band gap energy. This is similar to behavior observed for other devices (see, e.g., Table 2 of \citealt{2014PASP..126..739R}). Interestingly, at 500 nm or 0.6 eV above threshold, the probability of collecting 2 holes is only $\sim 4\%$, which is much less than what we would expect from studies of other semiconductors such as Si, Ge, or 5 $\mu$m cutoff HgCdTe \citep[Fig.~7]{2014PASP..126..739R} --- the {\slshape JWST} NIRSpec detectors, for example, show a $\sim 30\%$ probability at $E-E_{\rm t}=0.6$ eV. Regardless of the microphysical cause, the low probability of collecting 2 holes reduces the signal:noise ratio of the charge diffusion measurement. Another unexpected behavior that we observe is a peak at $\lambda^{-1} \sim 1.8\,\mu$m$^{-1}$ ($\lambda \approx 560$ nm), followed by a decline as we go to the blue. Note that there is good agreement with our analysis of the flats + darks in \S\ref{ss:recover} at 500 nm. We discuss possible explanations for the declining quantum yield in \S\ref{sec:discussion}.

The 2-hole probabilities in Fig.~\ref{fig:omega_lambda} show an offset from zero at small wavenumber (where generation of a second electron-hole pair is not energetically possible). This offset ranges from 0.15\% for SCA 20663 to 0.47\% for SCA 20828; such an offset could arise if there were a systematic difference in gain between the quantum efficiency and full well flat fields. It is worth noting that the full well data reaches saturation, and that of the three detector arrays studied here, SCA 20663 had the lowest persistence and burn-in and SCA 20828 had the highest \citep{2020PASP..132g4504F}. We suspect that burn-in effects in the infrared flat field data (the full well data) may be affecting the gain determination and contributing to the offset. We expect we could reduce this problem by acquiring additional flat field data that cover the required range of signal level, but does not approach saturation.

\subsection{Impact of charge diffusion on the shear signal}

\begin{figure}
    \centering
    \includegraphics[width=0.7\textwidth]{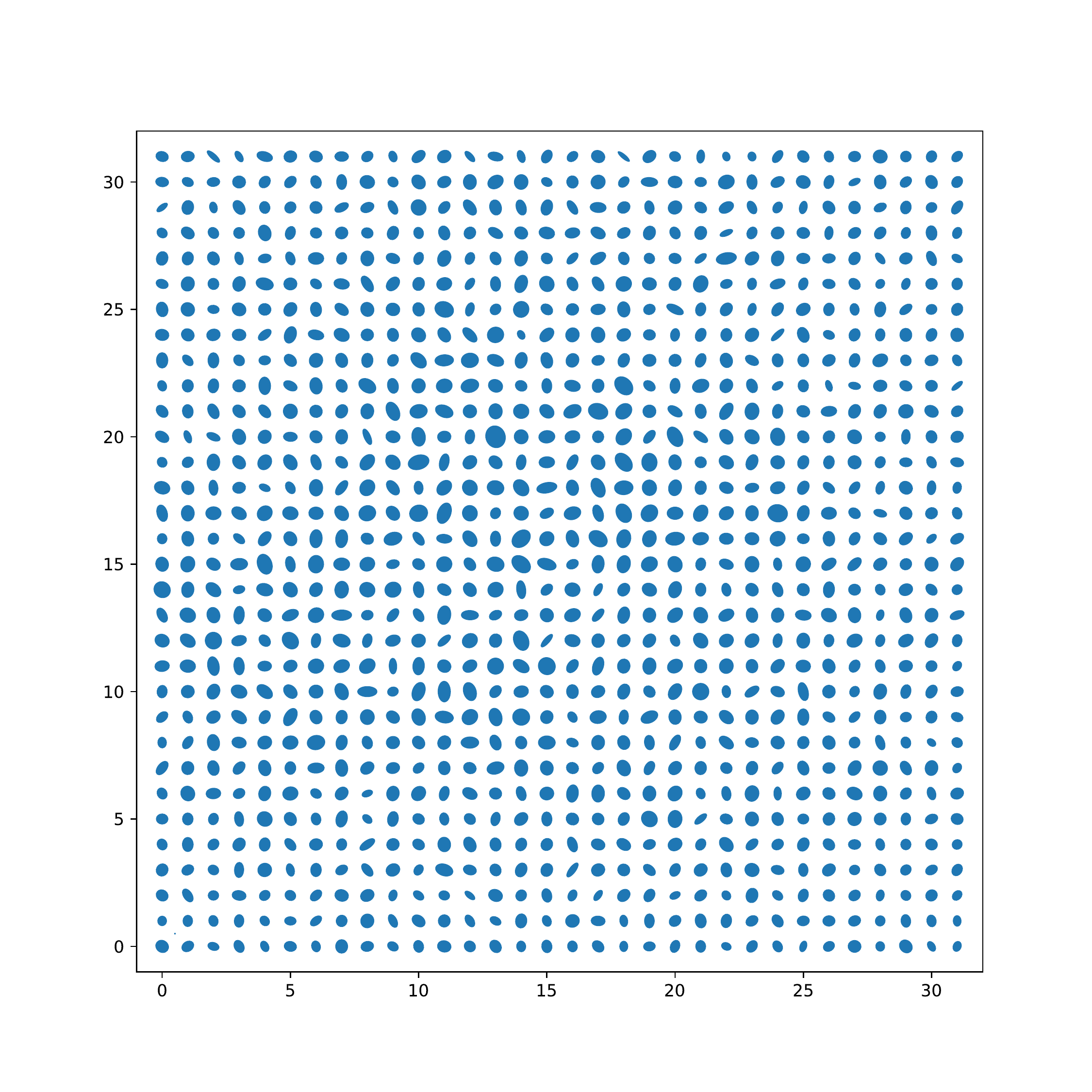}
    \caption{Elliptical grid based on charge diffusion covariance matrices for SCA 20829. The radius of each ellipse is set to two standard deviations.  The vertical and horizontal axes label superpixel indices.}
    \label{fig:elliptical_field}
\end{figure}

\begin{figure}
    \centering
    \includegraphics[width=0.5\textwidth]{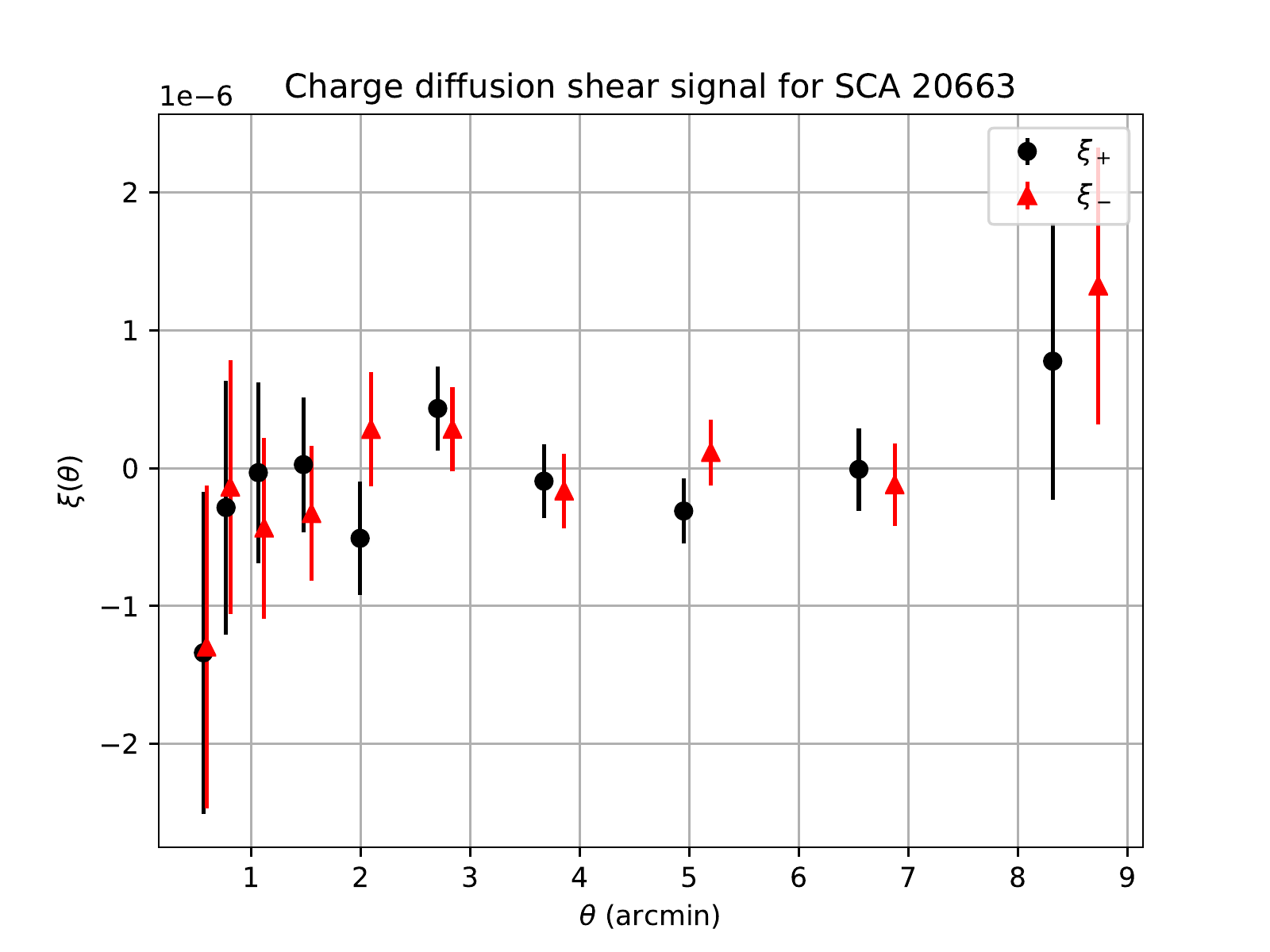}
    \includegraphics[width=0.5\textwidth]{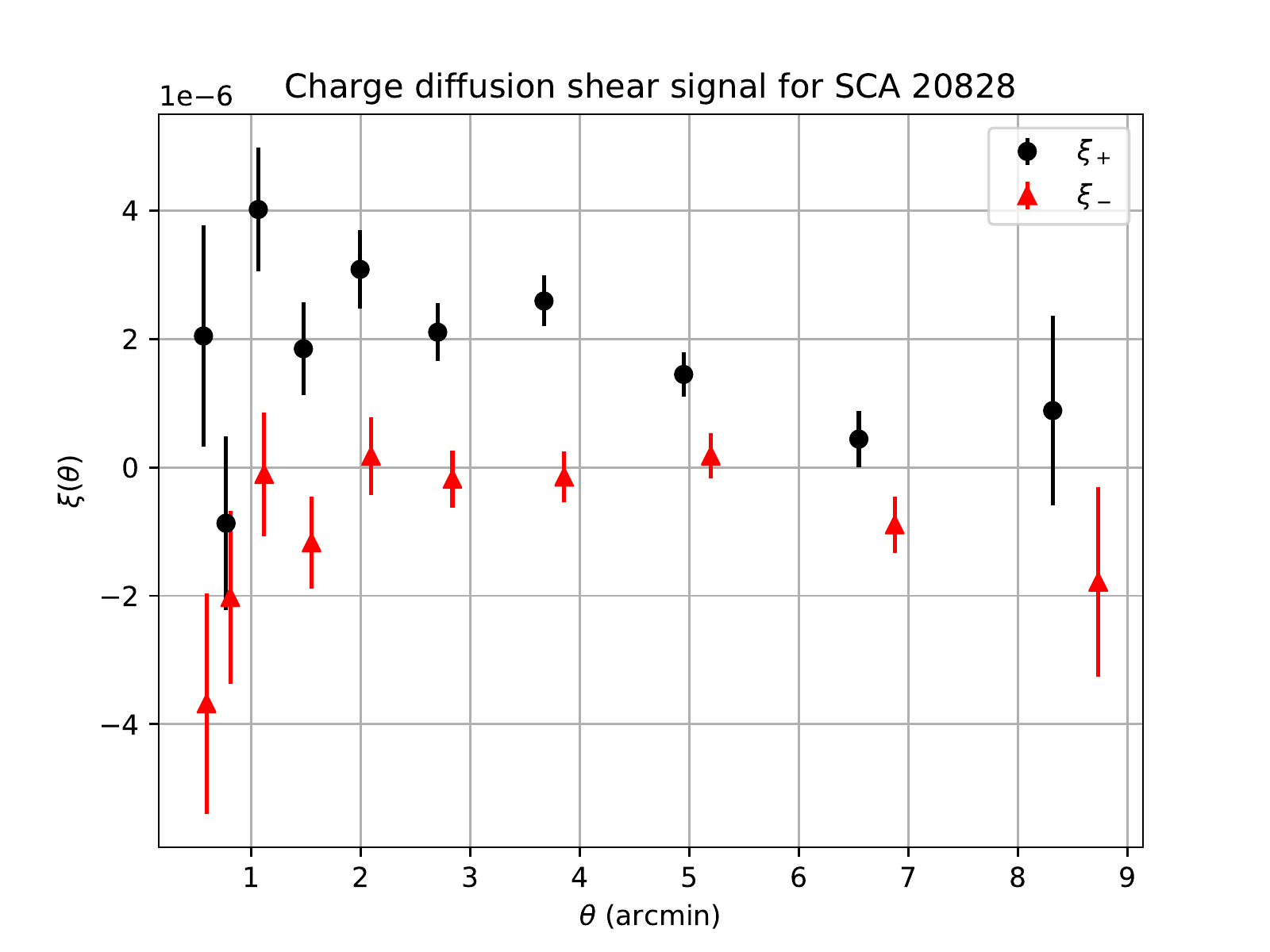}
    \includegraphics[width=0.5\textwidth]{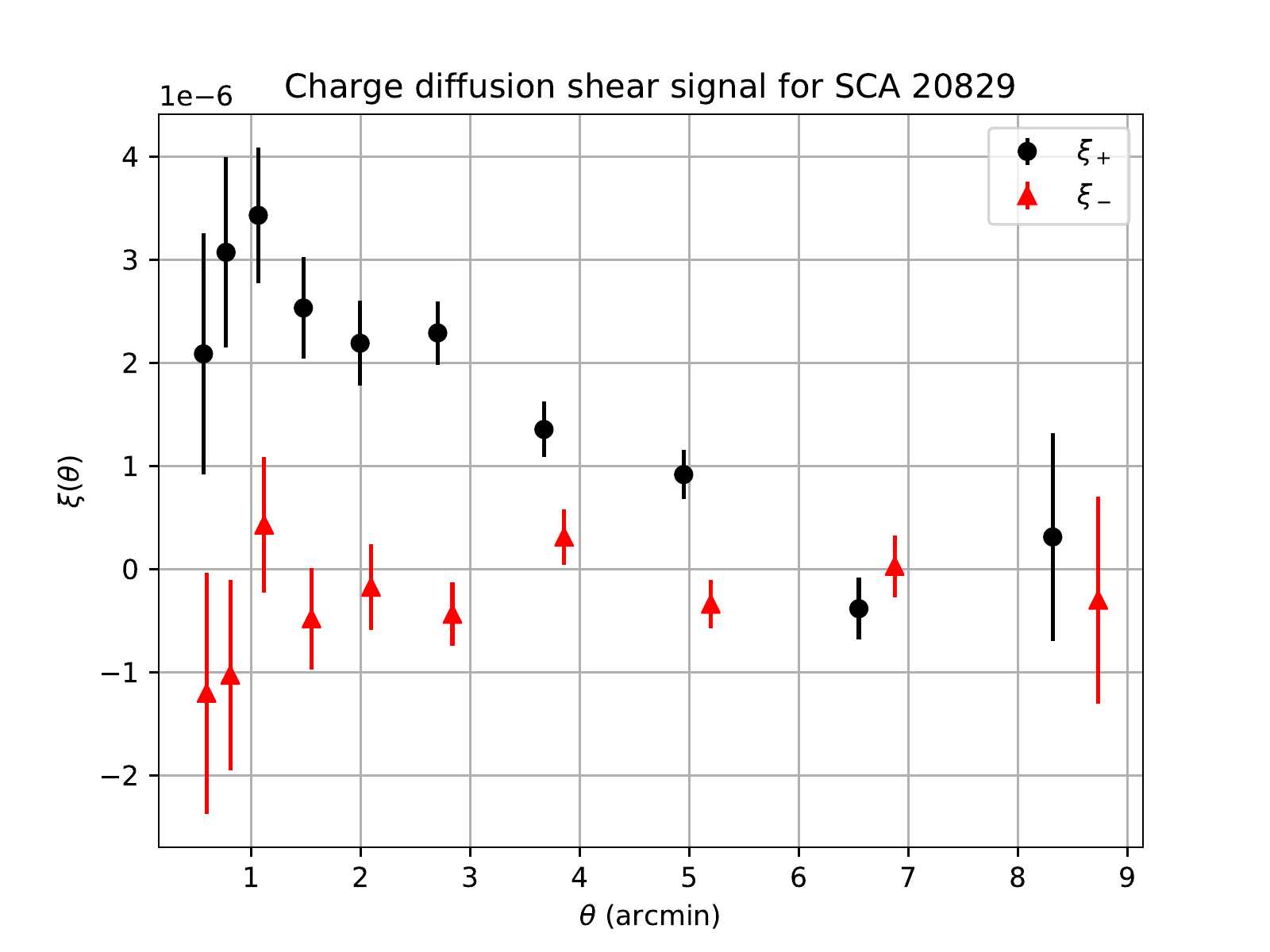}
    \caption{\label{fig:Xi20663}{\em Top panel:} Charge diffusion correlation functions $\xi_{+}(\theta)$ (black circles) and $\xi_{-}(\theta)$ (red triangles) measured on SCA 20663 data by {\sc TreeCorr}. A 5\% offset in $\theta$ was added to the plot of $\xi_-$ for ease of visualization. {\em Middle panel:} Same for SCA 20282 (the three super pixels where {\sc Solid-waffle} failed to converge are masked). {\em Bottom panel:} Same for SCA 20829.}
\end{figure}

To estimate the impact of charge diffusion on the weak lensing shear signal, we take a fiducial set of visible detector runs --- whose details are outlined in Table \ref{tab:config_vis_fid} --- and measure their respective charge diffusion covariance matrices (see \S\ref{sec:Gauss}) in $32 \times 32$ bins of superpixels. We visualize the covariance matrix in each superpixel using publicly available code\footnote{Specifically, we use the {\tt plot\_cov\_ellipse} function available from the BurnMan team at \newline {\tt https://github.com/geodynamics/burnman/blob/master/burnman/nonlinear\_fitting.py}} which prints an ellipse associated with each covariance. This is shown for SCA 20829 in Figure \ref{fig:elliptical_field}. Given an ellipticity field for each detector (and the fact that this work is a weak lensing analysis), it is natural for us to consider converting these shapes into a set of shears just as one would do for a galaxy ellipticity field on a patch of sky.
The conversion from components of charge diffusion covariance to shear can be obtained using the methodology in \citet{2021MNRAS.501.2044T}. There the aim was to estimate the effect of image motion on the galaxy ellipticity and hence the shear, but since both image motion and charge diffusion convolve the point spread function (PSF) with a kernel --- either the probability density function of the image motion in that case or $P_{\rm cd}({\boldsymbol\xi})$ here --- the results from \citet{2021MNRAS.501.2044T} can be carried over to our case. In the worst-affected redshift bin, the shear error induced by charge diffusion is
\begin{equation}
\begin{aligned}
    &\gamma_1=K_{\theta\theta}(C_{11}-C_{22}), \\
    &\gamma_2=2K_{\theta\theta}C_{12},    
\end{aligned}
\end{equation}
where $K_{\theta\theta}=1.432\times10^{-5}\textrm{~mas}^{-2}=0.1733\textrm{~pix}^{-2}$ \citep[Table B1]{2021MNRAS.501.2044T}. Next we pass a list of superpixel indices (converted to arcsec) and shears to {\sc TreeCorr}\footnote{https://github.com/rmjarvis/TreeCorr} \citep{2004MNRAS.352..338J} which calculates the shear correlation functions $\xi_{+}(\theta)$ and $\xi_{-}(\theta)$. Results for the three SCAs are shown in Figure~\ref{fig:Xi20663}.

{\slshape Roman} requirements on additive shear variance per component in each bin of multipole $\ell$ are given in its Science Requirements Document. A single detector has maximum and minimum separations corresponding to harmonic scales $10^{3}<\ell<10^{3.5}$ ($0.001 < \theta \textrm{ (rads)} <0.0031$)\footnote{The true minimum configuration space separation allowed is one superpixel or $\sim 66$ $\mu$rad. We quote a larger ``minimum'' separation in this subsection based on bins given in {\slshape Roman} requirements. The maximum quoted here is the diagonal of a single detector.} which lies in the highest $\ell$ bin. We compare the charge diffusion ``shear'' to {\slshape Roman} requirements using three different approaches, one in configuration space and two in harmonic space. This is done as a consistency check since each approach uses different approximations; once all 18 flight detectors are selected and their layout in the focal plane is chosen, it will be possible to tile the sky and unambiguously predict the spurious shear due to charge diffusion for a given tiling strategy.

In the first approach, we start with the equation
\begin{equation}
    \xi_+(\theta) = \int_0^\infty \frac{\ell d\ell}{2\pi} J_0(\ell\theta)[C_{EE}(\ell)+C_{BB}(\ell)] 
    \simeq \sum_{\ell \textrm{ bins}} 2\gamma^2 \overline{J_0(\ell\theta)},
\end{equation}
where $J_0$ is the Bessel function, $C(\ell)$ is the shear power spectrum (with the subscript 'EE' or 'BB' referring to the E-mode and B-mode power spectra components, respectively), $\gamma$ is the component of shear, and the overline indicates the average value over $\ell$. In the $\ell$ bin of interest, $\gamma=1.9\times10^{-4}$ and we make the rough estimate $\overline{J_0(\ell\theta)} \sim \mathcal{O}(10^{-1})$ for ease of calculation. The ratio of the measured charge diffusion correlation function to our estimate of {\slshape Roman} requirements is $\xi_{+,\textrm{ Fig.9}}/\xi_{+,\textrm{ Eq.36}} \sim 20$ over smaller separations in the detector.

In the second approach, we take the $\xi_+$ correlation function measured by {\sc TreeCorr} and Hankel transform it using the {\tt hankel} Python library \citep{2019JOSS....4.1397M} to get $C_{EE}(\ell)+C_{BB}(\ell)$. The variance per component over the $\ell$ bin for SCA 20829 is 
\begin{equation}\label{eq:var}
    \textrm{var per comp} = \int_{10^3}^{10^{3.5}}\frac{d\ell}{4\pi}\ell[C_{EE}(\ell)+C_{BB}(\ell)] = -2.2\times10^{-6}.
\end{equation}
This is $|\textrm{var per comp}/\gamma^2| \sim 60$ greater than requirements.

For the final comparison, we obtain theory predictions for $\xi_+$ from the publicly available theory pipeline {\tt CosmoSIS} \citep{2015A&C....12...45Z}, artificially tweaking $\sigma_8$ to get theory predictions that roughly agree with the measured values of $\xi_+$ values for each detector. {\tt CosmoSIS} gives the associated $C_\ell$'s which we plug into Eq.~(\ref{eq:var}). The variance per component calculated this way is $4.5\times 10^{-7}$, which is a factor of $\sim 12$ above requirements.

The above results show factors of systematic error in power which are all $\mathcal{O}(10)$. Weak lensing systematics mitigation through point spread function fitting acts on amplitudes. This means that systematic errors on shear need to be reduced by $\mathcal{O}(1)$ (i.e., the square root of the systematic error in power) which is doable using standard techniques.
Charge diffusion affects images of stars as well, so if it were perfectly linear (i.e., the covariance matrix of the charge diffusion is independent of the amount of charge already in the well), then it would be absorbed as part of PSF fitting. Next-order effects, such as wavelength and signal dependence of charge diffusion, could be expected since the conversion depth depends on the wavelength of illumination and the previously accumulated signal changes the electric field geometry (as evidenced by the BFE). The strength of these effects will determine how much complexity is needed in modeling and mitigating the systematics from charge diffusion.

\section{Discussion}\label{sec:discussion}
This paper has presented a formalism for the inclusion of quantum yield and charge diffusion effects in Fourier-space correlations of HxRG detector array visible flat and dark fields, expanding on the work of \citet{2020PASP..132g4504F} which considered only IR flat and dark field correlations. We updated {\sc Solid-waffle} to incorporate these changes and return measurements of visible characterization parameters in addition to the IR characterization parameters it previously output. We tested this updated machinery on simulations, finding we could recover IR and visible IPNL kernel parameters to within a few percent. 

We have also run our updated analyses on {\slshape Roman} flight candidate detectors SCAs 20663, 20828, and 20829 and presented the detailed results in \S\ref{sec:results}. For the fiducial set of configuration parameters, SCA 20663 had the largest values for the visible IPNL kernel. This is in contrast to the IR case where SCA 20828 holds this distinction. In all three detectors, both IR and visible IPNL kernel values are seen to be slightly asymmetric between rows and columns. For example, $([K^2a+KK^I]_{\rm H}-[K^2a+KK^I]_{\rm V})/([K^2a+KK^I]_{\rm H}+[K^2a+KK^I]_{\rm V}) = -0.052 \pm 0.005$ (SCA 20829 IR) and $([K^2a+KK^I]_{\rm H}-[K^2a+KK^I]_{\rm V})/([K^2a+KK^I]_{\rm H}+[K^2a+KK^I]_{\rm V}) = -0.065 \pm 0.007$ (SCA 20829 vis). The asymmetry for the IR and visible data are fairly consistent at $\sim 1.5\sigma$.

All detectors showed statistically significant amounts of quantum yield and charge diffusion at 500 nm. The quantum yield measured in each SCA begins to increase at $\lambda^{-1} > 1.2\, \mu$m$^{-1}$, which is similar to what we would expect scaling from 5 $\mu$m cutoff H2RGs \citep{2014PASP..126..739R}. This is expected behavior since in both cases a minimum energy is needed to produce the second electron-hole pair. However, at 500 nm, the quantum yield measured in this paper ($\omega \sim 3.5-3.9 \%$) is smaller than the fits obtained by \citet{2014PASP..126..739R}; their fit implies a quantum yield (our $1+\omega$) of 1.65 at 600 nm and 1.27 at 1 $\mu$m (photon energy of $5\times$ the band gap energy, as for {\slshape Roman} detectors at 500 nm).
We do not necessarily expect the production of the second electron-hole pair to simply scale with the band gap, however, since most of the band structure does not simply scale.\footnote{For example, the $E_1$ critical point of 2.5 $\mu$m cutoff HgCdTe is at $1/\lambda = E_1/hc=2.0$ $\mu$m$^{-1}$ and dominates absorption of 500 nm light \citep{1999JAP....85.2854D}, whereas for 5 $\mu$m cutoff HgCdTe this critical point only shifts down to 1.9 $\mu$m$^{-1}$ \citep{1984PhRvB..29.6752V}.}
 
Another possible explanation is that a significant fraction of the photons are getting absorbed in the buffer layer, as shown in \citet[][Fig.~3]{2020JATIS...6d6001M}, where the band gap is wider and the probability of a second electron-hole pair is less than in a $2.5 \, \mu$m band gap material. Our results from \S\ref{sec:qy vs wavelength}, in which the quantum yield peaks at $\sim 560$ nm and then declines at shorter wavelengths, also seem plausible in this picture, since the mean free path is very short at those wavelengths \citep[see][Fig.~7]{1999JAP....85.2854D}.

The charge diffusion $\sigma_{\rm cd} $ of 2.7--3.5 $\mu$m measured in this paper is larger than the prediction of 2.04 $\mu$m from the random walk model \citet[][Eq.~22]{2020JATIS...6d6001M}. It is possible that the numerical pre-factor of $1/\pi$ in the \citet{2020JATIS...6d6001M} may be modified with a full treatment of the drift-diffusion problem (with, for example, a distribution of drift times), or that some of the input parameters are slightly different from the as-built values. More interestingly, \citet{2020JATIS...6d6001M} note that their $\sigma_{\rm cd} = 2.0\,\mu$m prediction is consistent with $^{55}$Fe X-ray events (in which there is an additional cross-talk to the nearest neighbor pixels beyond the IPC measured from the single pixel reset method). One potential source of this difference is where the X-rays are absorbed relative to the 500 nm visible photons. The X-rays produced by the $^{55}$Fe source are predominately Mn KL$_2$+KL$_3$ X-rays with energy 5.89 keV \citep{2003RvMP...75...35D, nistxrayenergies}, which in Hg$_{1-x}$Cd$_x$Te with $x=0.445$ have an opacity of 523 cm$^2$/g \citep{nistxcom}; at the absorber density of $\rho = 7.07$ g/cm$^3$ \citep{capper94}, this corresponds to an absorption depth of $(\kappa\rho)^{-1} = 2.7\,\mu$m, or about half of the thickness of the Hg$_{1-x}$Cd$_x$Te absorber. In contrast, the absorption depth of a photon at 500 nm in Hg$_{1-x}$Cd$_x$Te is 28 nm (\citealt{1999JAP....85.2854D}; in practice, such a photon may be absorbed in the buffer layer). We thus anticipate that holes released by X-ray absorption events likely travel only part way through the absorbing layer, and thus on average may give a lower charge diffusion length than a measurement with visible light.

This work made the first estimate of how much the measured charge diffusion should affect the weak lensing shear signal. For all detectors, we find the charge diffusion shear correlation function $\xi_+ \sim 10^{-6}$ which is about 1\% of the galaxy shear correlation function. We compared the measured charge diffusion $\xi_+$ (and $C_\ell$) for SCA 20829 to {\slshape Roman} requirements on additive shear per component in the highest $\ell$ bin and found measured values that are $\mathcal{O}(10)$ larger. A more detailed analysis of this effect will be carried out once the layout of all 18 detectors across the {\slshape Roman} focal plane is chosen. Note that this impact is described without taking into account any PSF mitigation. The standard approach taken in weak lensing analyses is to determine a PSF based on stars and interpolate between star positions on the focal plane \citep[for a review, see Section 2.1 of][]{2018ARA&A..56..393M}. In the case where the density of stars is not enough for direct interpolation (e.g., with a polynomial fit) to capture PSF variations on the smallest spatial scales, principal component methods can be applied to use all available data to identify the most important dimensions of PSF variance \citep{2004astro.ph.12234J} and have been shown to be effective on space-based data from the Hubble Space Telescope Advanced Camera for Surveys \citep{2007PASP..119.1403J,2010A&A...516A..63S}. A potential choice to investigate in future work would be to include the 500 nm charge diffusion map or a similar map from the in-flight $r$-band flat fields as an additional basis mode to constrain the PSF fit.

This analysis is one of many that are required to calibrate effects present in {\slshape Roman} detectors. Such calibration will be necessary for the accurate determination of scientific results, including cosmological parameter estimates, from {\slshape Roman} pipelines once the survey begins collecting data. We have extended the correlation function formalism of \citet{2020PASP..132g4504F} to account for quantum yield and charge diffusion effects in the presence of visible light. We applied this updated formalism to three {\slshape Roman} flight candidates --- SCAs 20663, 20828, and 20829 --- and measured over a dozen new parameters. Just as we did in our previous investigation \citep{2020PASP..132g4504F}, we caution readers that all data were taken under laboratory conditions and that there will likely be some differences with {\slshape Roman} observations, most notably the flight controller (ACADIA, \citealt{2018SPIE10709E..0TL}) instead of the Leach controller. In future work we will carry out the same analyses done in this paper on other flight candidates, and with data acquired with the flight SCAs connected to an ACADIA controller. We also plan to compare the charge diffusion measurements obtained via the visible flat field method to measurements obtained with spot projection (\citealt{2013PASP..125.1065S}; a similar test is planned for one of the {\slshape Roman} development detectors). The two methods are expected to have complementary strengths: the visible flat field method provides true averages over regions on the array, and only corresponds to diffusion of charge without the optical PSF or image motion; on the other hand, the spot projection is more similar to what happens when one observes a PSF star, and can be used over the full range of wavelengths including in the NIR where multiple electron-hole pair production does not occur.

{\em Software:} Astropy \citep{2013A&A...558A..33A,2018AJ....156..123A}, fitsio \citep{fitsio}, Matplotlib \citep{Hunter:2007}, NumPy \citep{numpy}, SciPy \citep{2020SciPy-NMeth}

\section*{Acknowledgements}

We thank Jeff Kruk and Charles Shapiro for helpful comments on the draft of this paper.

This work was supported by contract NASA 15-WFIRST15-0008. C.H. is also supported by the Simons Foundation and the David \& Lucile Packard Foundation.

\appendix

\section{Frames and configuration file passed to {\sc Solid-waffle}}
Here we provide the sequence of frames (including their full location within the Ohio Supercomputer Center) passed to {\sc Solid-waffle} via a configuration file for each detector. These frames are paired with the settings given in Table \ref{tab:config_vis_fid}. Tables \ref{frames_20663}, \ref{frames_20828}, and \ref{frames_20829} correspond to SCAs 20663, 20828, and 20829, respectively. These frames are paired with the settings given in Table \ref{tab:config_vis_fid}.

\begin{table}[]
    \centering
    \begin{tabular}{l}
    \hline
     LIGHT: \\
     $\sim$/SCA20663/20190926\_95K\_1p1m0p1\_ch1\_1400nm\_gr3\_filt5\_shutter\_open\_20663\_001.fits \\
     \vdots \; (increment last value by 001) \\
     $\sim$/SCA20663/20190926\_95K\_1p1m0p1\_ch1\_1400nm\_gr3\_filt5\_shutter\_open\_20663\_005.fits \\
     $\sim$/SCA20663/20190926\_95K\_1p1m0p1\_ch3\_1400nm\_gr3\_filt5\_shutter\_open\_20663\_001.fits \\
     \vdots \; (increment last value by 001) \\
     $\sim$/SCA20663/20190926\_95K\_1p1m0p1\_ch3\_1400nm\_gr3\_filt5\_shutter\_open\_20663\_011.fits \\
     DARK: \\
     $\sim$/SCA20663/20190926\_95K\_1p1m0p1\_ch0\_1400nm\_gr3\_filt5\_shutter\_closed\_20663\_001.fits \\
     \vdots \; (increment last value by 001) \\
     $\sim$/SCA20663/20190926\_95K\_1p1m0p1\_ch0\_1400nm\_gr3\_filt5\_shutter\_closed\_20663\_010.fits \\
     $\sim$/SCA20663/20190926\_95K\_1p1m0p1\_ch2\_1400nm\_gr3\_filt5\_shutter\_closed\_20663\_002.fits \\
     \vdots \; (increment last value by 001) \\
     $\sim$/SCA20663/20190926\_95K\_1p1m0p1\_ch2\_1400nm\_gr3\_filt5\_shutter\_closed\_20663\_007.fits \\
     VISLIGHT: \\
     $\sim$/SCA20663/20190926\_95K\_1p1m0p1\_ch21\_500nm\_gr3\_filt6\_shutter\_open\_20663\_001.fits \\
    $\sim$/SCA20663/20190926\_95K\_1p1m0p1\_ch23\_500nm\_gr3\_filt6\_shutter\_open\_20663\_001.fits \\
    $\sim$/SCA20663/20190926\_95K\_1p1m0p1\_ch25\_500nm\_gr3\_filt6\_shutter\_open\_20663\_001.fits \\
    \vdots \; (repeat the set of three lines above twice, once ending with 002 and once with 003) \\
    VISDARK: \\
    $\sim$/SCA20663/20190926\_95K\_1p1m0p1\_ch20\_500nm\_gr3\_filt6\_shutter\_closed\_20663\_002.fits \\ \vdots \; (increment last value by 001) \\
    $\sim$/SCA20663/20190926\_95K\_1p1m0p1\_ch20\_500nm\_gr3\_filt6\_shutter\_closed\_20663\_010.fits \\ \hline

    \end{tabular}
    \caption{Frames passed to configuration file for the fiducial set up of SCA 20663.}
    \label{frames_20663}
\end{table}

\begin{table}
     \centering
     \begin{tabular}{l}
\hline
LIGHT: \\ 
 $\sim$/SCA20828/20191204\_95K\_1p1m0p1\_ch1\_1400nm\_gr3\_filt5\_shutter\_open\_20828\_001.fits \\ 
 \vdots \; (increment ch values by 2, i.e. go to the next odd value)\\
 $\sim$/SCA20828/20191204\_95K\_1p1m0p1\_ch17\_1400nm\_gr3\_filt5\_shutter\_open\_20828\_001.fits \\
 $\sim$/SCA20828/20191204\_95K\_1p1m0p1\_ch1\_1400nm\_gr3\_filt5\_shutter\_open\_20828\_002.fits \\  
 \vdots\; (increment ch values by 2, i.e. go to the next odd value)\\
  $\sim$/SCA20828/20191204\_95K\_1p1m0p1\_ch13\_1400nm\_gr3\_filt5\_shutter\_open\_20828\_002.fits \\ 
 DARK: \\ 
 $\sim$/SCA20828/20191204\_95K\_1p1m0p1\_ch0\_1400nm\_gr3\_filt5\_shutter\_closed\_20828\_001.fits \\ \vdots \; (increment last value by 001) \\
 $\sim$/SCA20828/20191204\_95K\_1p1m0p1\_ch0\_1400nm\_gr3\_filt5\_shutter\_closed\_20828\_009.fits \\
 $\sim$/SCA20828/20191204\_95K\_1p1m0p1\_ch2\_1400nm\_gr3\_filt5\_shutter\_closed\_20828\_002.fits \\
  \vdots \; (increment last value by 001) \\ 
 $\sim$/SCA20828/20191204\_95K\_1p1m0p1\_ch2\_1400nm\_gr3\_filt5\_shutter\_closed\_20828\_008.fits \\
 VISLIGHT: \\
 $\sim$/SCA20828/20191204\_95K\_1p1m0p1\_ch21\_500nm\_gr3\_filt6\_shutter\_open\_20828\_001.fits \\
 $\sim$/SCA20828/20191204\_95K\_1p1m0p1\_ch23\_500nm\_gr3\_filt6\_shutter\_open\_20828\_001.fits \\
 $\sim$/SCA20828/20191204\_95K\_1p1m0p1\_ch25\_500nm\_gr3\_filt6\_shutter\_open\_20828\_001.fits \\
 \vdots \; (repeat the set of three lines above twice, once ending with 002 and once with 003) \\ 
 VISDARK: \\
 $\sim$/SCA20828/20191204\_95K\_1p1m0p1\_ch20\_500nm\_gr3\_filt6\_shutter\_closed\_20828\_002.fits \\ \vdots \; (increment last value by 001) \\ $\sim$/SCA20828/20191204\_95K\_1p1m0p1\_ch20\_500nm\_gr3\_filt6\_shutter\_closed\_20828\_010.fits  \\ \hline
 \end{tabular}
 \caption{Frames passed to configuration file for the fiducial set up of SCA 20828.}
 \label{frames_20828}
 \end{table}

\begin{table}
     \centering
     \begin{tabular}{l}
\hline
LIGHT: \\ 
 $\sim$/SCA20829/20191018\_95K\_1p1m0p1\_ch1\_1400nm\_gr3\_filt5\_shutter\_open\_20829\_001.fits
 \\ \vdots \; (increment ch values by 2, i.e. go to the next odd value)\\
 $\sim$/SCA20829/20191018\_95K\_1p1m0p1\_ch17\_1400nm\_gr3\_filt5\_shutter\_open\_20829\_001.fits \\
 $\sim$/SCA20829/20191018\_95K\_1p1m0p1\_ch1\_1400nm\_gr3\_filt5\_shutter\_open\_20829\_002.fits \\
 \vdots\; (increment ch values by 2, i.e. go to the next odd value)\\
 $\sim$/SCA20829/20191018\_95K\_1p1m0p1\_ch13\_1400nm\_gr3\_filt5\_shutter\_open\_20829\_002.fits \\ 
 DARK: \\ 
 $\sim$/SCA20829/20191018\_95K\_1p1m0p1\_ch0\_1400nm\_gr3\_filt5\_shutter\_open\_20829\_001.fits \\ \vdots \; (increment last value by 001) \\
 $\sim$/SCA20829/20191018\_95K\_1p1m0p1\_ch0\_1400nm\_gr3\_filt5\_shutter\_open\_20829\_009.fits \\
 $\sim$/SCA20829/20191018\_95K\_1p1m0p1\_ch2\_1400nm\_gr3\_filt5\_shutter\_open\_20829\_002.fits \\
  \vdots \; (increment last value by 001) \\ 
 $\sim$/SCA20829/20191018\_95K\_1p1m0p1\_ch2\_1400nm\_gr3\_filt5\_shutter\_open\_20829\_008.fits \\
 VISLIGHT: \\
 $\sim$/SCA20829/20191018\_95K\_1p1m0p1\_ch21\_500nm\_gr3\_filt6\_shutter\_open\_20829\_001.fits \\
 $\sim$/SCA20829/20191018\_95K\_1p1m0p1\_ch23\_500nm\_gr3\_filt6\_shutter\_open\_20829\_001.fits \\
 $\sim$/SCA20829/20191018\_95K\_1p1m0p1\_ch25\_500nm\_gr3\_filt6\_shutter\_open\_20829\_001.fits \\
 \vdots \; (repeat the set of three lines above twice, once ending with 002 and once with 003) \\ VISDARK: \\
 $\sim$/SCA20829/20191018\_95K\_1p1m0p1\_ch20\_500nm\_gr3\_filt6\_shutter\_closed\_20829\_002.fits \\ \vdots \; (increment last value by 001) \\ $\sim$/SCA20829/20191018\_95K\_1p1m0p1\_ch20\_500nm\_gr3\_filt6\_shutter\_closed\_20829\_010.fits \\ \hline
 \end{tabular}
 \caption{Frames passed to configuration file for the fiducial set up of SCA 20829.}
 \label{frames_20829}
 \end{table}
 
 \begin{table}[]
    \centering
    \begin{tabular}{ll}
    \hline
    FORMAT     &  4 \\ \hline
    CHAR     &  Advanced 1 3 3 bfe \\ \hline
    TIMEREF & 0 \\ \hline
    NBIN & 32 32 \\ \hline
    FULLNL & True True True \\ \hline
    NLPOLY & 4 1 16 \\ \hline
    IPCSUB & True \\ \hline
    VISTIME & 4 51 1 3 \\ \hline
    VISBFETIME & 4 27 28 51 \\ \hline
    TIME & 1 8 9 16 \\ \hline
%    HOTPIX & 500 1000 0.1 0.1 \\ \hline
    MASK & 0 30 \\ \hline
    MASK & 0 31 \\ \hline
    MASK & 1 31 \\ \hline 
    \end{tabular}
    \caption{Configuration file settings used for the fiducial set of visible detector runs. All three flight candidates had the same inputs for FORMAT through TIME. Only SCA 20828 had masked pixels; other detectors need not have MASK passed as a setting. A description of each setting in the leftmost column can be found in {\sc Solid-waffle}'s {\tt ScriptInformation.txt} file.}
  \label{tab:config_vis_fid}
\end{table}
 
%\newpage

\bibliography{refs.bib}{}
\bibliographystyle{aasjournal}

\end{document}